\documentclass[preprint2]{aastex}
\usepackage{graphicx,amssymb,amstext,enumerate,amsmath}
\usepackage{color,graphicx}

\shorttitle{NGC 3516 ionized absorbers}
\shortauthors{Huerta E. M. et al.\ }

\begin{document}

\title{A DETAILED ANALYSIS OF THE HIGH RESOLUTION X-RAY SPECTRA OF NGC 3516: \\
         VARIABILITY OF THE IONIZED ABSORBERS }
\author{E. M. Huerta\altaffilmark{1,2}, Y. Krongold\altaffilmark{1}, F. Nicastro\altaffilmark{3}, S. Mathur \altaffilmark{4} , A. L. Longinotti\altaffilmark{5} \& E. Jimenez-Bailon\altaffilmark{1}.}
\altaffiltext{1}{Departamento de Astronomia Extragalactica y Cosmologia, Instituto de Astronomia, Universidad Nacional
  Autonoma de Mexico (UNAM), Apartado Postal 70-264, 04510, Mexico DF, Mexico.}  
\altaffiltext{2}{Departamento de Fisica, Universidad Autonoma Metropolitana Iztapalapa (UAM-I), Av. San Rafael Atlixco 186, 09340, Mexico, D.F. Mexico.}
\altaffiltext{3}{Osservatorio Astronomico di Roma-INAF, Via di Frascati 33, 00040, Monte Porzio Catone, RM, Italy.}
\altaffiltext{4}{Astronomy Department, The Ohio State University, Enarson Hall, 140 W. 18th Avenue, Columbus, OH 43210, USA.}
\altaffiltext{5}{European Space Astronomy Centre of ESA, Madrid, Spain}

\begin{abstract}

The 1.5 Seyfert galaxy NGC 3516 presents a strong time variability in X-rays. We re-analyzed the 9 observations performed on October 2006 by XMM-Newton and 
Chandra in the 0.3 to 10 keV energy band. An acceptable model was found to the XMM-Newton data fitting EPIC-PN and RGS spectra simultaneously; 
later, this model was successfully applied to the  contemporary Chandra high resolution data. The model consists of a continuum emission component (power law + 
blackbody) absorbed by four ionized components (warm absorbers), and ten narrow emission lines. Three absorbing components are warm, producing features 
only in the soft X-ray band. The fourth ionization component produces FeXXV and FeXXVI in the hard-energy band.  
We study the time response of the absorbing components to the well detect changes in the X-ray luminosity of this source, and find that the two 
components with the lower ionization state show clear opacity changes consistent with gas close to photoionization equilibrium. These changes are supported by the models 
and by differences in the spectral features among the nine observations. On the other hand, the two components with higher ionization state do not seem to respond to 
continuum variations. The response time of the ionized absorbers allows us to constrain their electron density and location. We find that one component (with intermediate 
ionization) must be located within the obscuring torus, at a distance 2.7$\times$10$^{17}$ cm from the central engine. This outflowing component is likely originated in the 
accretion disk. The three remaining components are at distances larger than $10^{16-17}$ cm. Two of the  absorbing components in the soft X-rays have similar outflow 
velocities and locations. These components may be in pressure equilibrium, forming a multi-phase medium, if the gas has metallicity larger than the solar one (Z$\gtrsim 5 Z_{\odot}$). 
We also search for variations in the covering factor of the ionized absorbers (although partial covering is not required in our models). We find no correlation between the 
change in covering factor and the flux of the source. This, in connection with the observed variability of the ionized absorbers, suggests that the changes in flux are not 
produced by this material. If the variations are indeed produced by obscuring clumps of gas, these must be located much closer in to the central source.

\end{abstract}


\section{Introduction}

In $\sim50\%$ type 1 Active Galactic Nuclei ({\it AGN}) (e.g. \cite{geo98,pic05}) absorption lines are detected in the far UV and the soft X-ray spectral ranges 
\citep{mat95,mat97}. These lines, which are blue-shifted, are generated by several ionized species. Nearly one hundred features are produced in these media with outflow 
velocities of the order of few hundred  to thousand kms$^{-1}$ \citep{cren03,kro03}. The location of this absorbing material, referred to as the warm absorber 
({\it WA}) \citep{hal84} remains uncertain. Suggestions range from the accretion disk itself (\citealp{elv00}; Krongold et al. 2005a, 2007, 2010; \citealp{lon13}) to the obscuring 
torus \citep{blu05} and the narrow-line region (e.g. \cite{ric10}). There are several models that attempt to explain the distribution and physical conditions of the absorbing 
gas. One of them proposes that different WA phases can coexist in pressure equilibrium, a multiphase medium (Krongold et al. 2003, 2005b; \citealp{car09}). Another one 
suggests that the material, also in pressure equilibrium, is distributed with a radial stratification \citep{roz06}. Most X-ray analysis have been based on photoionization models 
(e.g. \cite{net03,mck07,and10,lon10}), where the properties of the absorbing gas are described in terms of  the Hydrogen equivalent column 
density and its ionization state. These studies have yielded WA multiphase winds with equivalent column density values N$_{H} \sim 10^{21-23}$ cm$^{-2}$. In this kind of 
models, the ionization state of the WA components is characterized by the ionization parameter (here defined as $U = \frac {Q}{4{\pi}R^{2}n_{H}c}$, where $Q$ is the rate of 
0.013-100 keV photons produced in the source, $R$ the distance between the black hole and the accretion disk system to WA location, $n_{H}$ the Hydrogen number density 
and c the light speed). As noted by \cite{nic99} and \cite{kro07} there is a degeneracy in this formula, as only the product $n_{e}R^{2}$ can be inferred from observables. 
However, time variability analysis solve this problem: by measuring the response time of the absorber to variations in the impinging continuum, the density $n_{H}$ can be 
estimated \citep{nic99} and therefore the distance from the source to the ionized absorber can be constrained (see also \cite{krol95,rey95}; Krongold et al. (2007, 2010)). 

Studying the properties of the absorbing material is of great importance, as it has been suggested that these outflows can play an important role in galaxy evolution. If the 
outflowing material is a significant fraction of the accreted mass they can provide the feedback proposed by the cosmological models \citep{matteo05,hop06}). 

\subsection{Seyfert 1.5 galaxy NGC 3516}

NGC 3516 is a Seyfert 1.5 galaxy \citep{ver06} at $z=0.00886$ \citep{keel96} that presents an extreme X-ray flux variability \citep{ede00,net02,tur08,mar08}. NGC 3516 is 
also extremely variable in optical and UV emission (see as examples: \cite{voi87,wal90,cre98,goa99,ede99,mao02,kra02}). 

The source has been observed in X-rays from 1979 \citep{mac87} to 2009 \citep{tur11} with all the X-ray observatories \citep{kol93,nan94,mor95,kri96b,ede99,cos00,gua01,
geo02,tur08,tur11}. NGC 3516 presents a complex spectrum: 
in the soft X-ray range (0.3 - 2 keV), the spectrum shows the so-called {\it Soft Excess}, several absorption features attributed to the presence of warm absorbers and 
few emission lines \citep{cos00,gua01,geo02,tur05,tur08,meh10}. In the hard band (2 to 200 keV), it presents the fluorescence Iron emission line K$\alpha$ and two 
absorption lines by FeXXV and FeXXVI. To fit the hard band X-ray continuum of this source \cite{tur05} fitted a Power Law, finding a Photon Index ${\Gamma}\sim$ 2 in the 
10 - 200 keV regime. Also \cite{mar08} fitted an absorbed Power Law with ${\Gamma}=$ 2 over the 0.3 - 76 keV range. Recently, \cite{meh10} reported a significant change in 
the slope of the Power Law among 2006 XMM-Newton observations: ${\Gamma}$ varied from 1.70$\pm$0.2 to 1.85$\pm$0.2, in addition to significant changes in the 0.2 to 
10 keV range flux.

To model the Soft Excess emission \cite{mar08} added a second Power Law, with the same Photon Index as the primary power law but not obscured by any absorber. 
\cite{meh10} added a modified Black Body component which describes the spectrum of a Black Body reprocessed by coherent Compton scattering. The best fit temperature 
varied from 186$\pm$3 eV to 211$\pm$4 eV. We note that the Soft Excess physical interpretation is still an issue (see \cite{ric10}). 

\subsection{Previous analysis of the Ionized Absorber in NGC 3516 \label{ngc3516}}

The complex absorption in NGC 3516 soft X-ray band was detected by two strong absorption features at 0.74 and 0.87 keV with {\it ROSAT} \citep{mat97} and 
{\it ASCA} X-ray telescopes \citep{kri96b}. These features were originally interpreted as absorbing K edges due to OVII and OVIII bound-free transitions. 
More recently, a few high ionization absorption components producing both absorption lines and edges have been detected in high resolution spectra. 
\cite{tur05} reported three ionization states: (1) a ``cool" absorber with a column density N$_{H}\sim$6$\times$10$^{21}$ cm$^{-2}$ and an outflow velocity v$_{out}$=-200 
km s$^{-1}$ (all the outflow velocities in this work were fixed to a given value), associated with the ``UV absorber'' found by \cite{kra02}; 
(2) a ``high ionization" absorber with N$_{H}\sim$10$^{22}$ cm$^{-2}$ and a fixed v$_{out}$=-1100 kms$^{-1}$; and (3) a ``heavy'' 
absorber with N$_{H}\sim$10$^{23}$ cm$^{-2}$ and v$_{out}$=-1100 kms$^{-1}$, that covers only 50\% of the continuum source. Later on, 
\cite{tur08} discovered a fourth absorbing component, with even higher ionization state, through the positive detection of H-like and He-like features of Fe XXV and XXVI. 
This component, only detectable in the hard X-ray  band, has  N$_{H}\sim$10$^{23}$ cm$^{-2}$ and v$_{out}\sim$-1000 kms$^{-1}$. 
\cite{tur08} interpreted the flux variations in the X-ray band as the result of variations in the covering fraction of the ``heavy absorber".

Using low resolution {\it Suzaku} data, \cite{mar06} found two absorbing components: A ``primary absorber" with properties consistent with the ``heavy'' 
absorber reported by \cite{tur08} (although with a covering fraction of 96 -100\%) and a mildly-ionized medium, consistent with the ``UV absorber''.  
\cite{meh10} analyzed the XMM Newton data of the source. They detected the same three absorbing components as \cite{tur08}, but they reported that partial covering is not 
required by the data for the heavy absorber. If these authors impose a partial absorber in their models, they further find that variations in the covering fraction are not 
consistent with the data (in contrast to the results by \cite{tur08}). 

Finally, \cite{hol12} found a kinematic structure in the absorber, consisting in four different outflow components intrinsic to NGC 3516: (1) 
v$_{out}$=-350 $\pm$ 100 kms$^{-1}$, (2) v$_{out}$=-1500$\pm$150 kms$^{-1}$, (3) v$_{out}$=-€"2600$\pm$200 kms$^{-ˆ'1}$, and (4) 
$v_{out}$=-4000$\pm$400 kms$^{-1}$. \cite{hol12} analyzed the Chandra high resolution spectra obtained in 2001 and 2006. They detect variability in this 
structure, noting that the high velocity components are only present in the 2006 observations. They further report that the covering factor plays a 
minor role in the absorption lines.

In the UV band \cite{kra02} have reported eight different kinematic absorbing components. Using data obtained by the {\it GHRS} (Goddard High Resolution 
Spectrograph) on board the Hubble Space Telescope ({\it HST}), these authors detected four of these components through  Ly${\alpha}$,  CIV, and NV absorption lines with 
radial velocities of -376, -183 and -36 kms$^{-1}$ (where the last system is comprized by a blend of two different kinematic components). Later on, the other four kinematic 
components were detected with the Space Telescope Imaging Spectrograph ({\it STIS}) on board the HST. These components present radial velocities of -692, -837, 
-994, and -1372 kms$^{-1}$. These authors suggest that the emergence of the new components is the result of variations of ionized gas in response to changes in the
ionizing continuum. 

\subsection{X-ray Variability  on NGC 3516}

As it has been reported NGC 3516 presents strong flux variations in time, from the optical-UV to the X-rays (see for example \cite{kor96} and \cite{kra02}). The 
source presents extreme X-ray flux variations, by a factor of 5 on timescales of hours \citep{tur08}, and by a factor of 50 in timescales of years \citep{net02}. 
These variations are often associated with spectral changes (e.g.  \cite{net02,tur08} and references therein) not only in the energy flux, there is evidence of spectral 
features variability \citep{mat97,net02,tur08,hol12}. To understand the nature of the variability, some scenarios invoking obscuration have been proposed: \cite{cos00} 
suggested neutral clouds that hide the central source (using {\it Beppo SAX} telescope spectra), \cite{mar08} (with Suzaku spectra) proposed a model in which flux variations 
could be explained by the presence of discrete blobs or filaments located within few light years of the black hole, traversing the line of sight during the observation time. 
\cite{tur08} suggested a simple explanation were that only parameter varying among observations (and producing both the flux and spectral variations) is the covering factor 
of the heavy absorber phase.

Other interpretations suggest that rather, the variations might be intrinsic to the source. \cite{meh10} arrived to this conclusion using 2006 XMM Newton spectra. 

However, in a most recent X-ray NGC 3516 analysis, using Suzaku 2009 X-ray observations, \cite{tur11} claims an absence of reverberation signals. This led them to
conclude that intrinsic continuum variability is not possible. They suggested that the variability can be a consequence of Compton-thick clumps of gas in the line-of-slight. 

In this paper we present the analysis of all 2006 X-ray observations available in the XMM-Newton Science Archive ({\it XSA}) and Chandra Data Archive ({\it CDA}) of 
NGC 3516. We study the both high and low resolution spectra to cover the range between 0.3 keV to 10 keV. We first present the analysis over individual spectra to 
characterize the spectral components, and then study the time variability of the absorber among observations. In our analysis we find variations in two of the ionized 
absorbers in response to continuum variations.

Section \S 2 describes the X-ray observations and the data reduction. In section \S 3, we present 2006 X-ray data analysis and the best model for XMM data, we 
also describe the Chandra analysis. Section \S 4 contains our results, including the time variability behaviour and a physical scenario proposed to explain the X-ray 
proprieties and the flux variability of NGC 3516. We discuss our results in section 5. Finally in section 6 we present the conclusions.

\section{X-ray Observations and Data Reduction}

X-ray observations of NGC 3516 were obtained with XMM-Newton \citep{jan01} and Chandra X-ray Observatory \citep{wei00} in October 2006. The log of observations is 
presented in Tables \ref{xmm} and \ref{cxo}, for XMM-Newton and Chandra X-ray observatories respectively. In the following we describe the data properties and reduction 
process.

\begin{deluxetable}{rrccccc} 
\rotate
\tabletypesize{\scriptsize}
\tablecolumns{7} 
\tablewidth{0pc} 
\tablecaption{Observations Mean Count Rate and Exposure time of XMM-Newton observations taken in October of 2006 \label{xmm}} 
\tablehead{ 
\colhead{\bf Observation} &  \colhead{ObsID}  & \colhead{Obs. Date} & 
\multicolumn{2}{c}{Exposure time} & \multicolumn{2}{c}{Mean Count Rate} \\
\colhead{} & \colhead{} & \colhead{[Year/Month/Day]} & \multicolumn{2}{c}{[seconds]}  & \multicolumn{2}{c}{[counts/sec] $\pm \sigma$}  \\
\colhead{} & \colhead{} & \colhead{} & \colhead{\bf PN} & \colhead{\bf RGS} & 
\colhead{\bf PN} & \colhead{\bf RGS} }
\startdata
1 XMM [{\bf 1x}] & $0401210401$ & 2006-10-06 & $36142$ & $51558$ & 30.80 & 0.90 \\ 
2 XMM [{\bf 2x}] & $0401210501$ &  2006-10-08  & $48031$ &  $68752$  & 28.60 & 0.84 \\ 
3 XMM [{\bf 5x}] & $0401210601$ &  2006-10-10  &  $47534$  &  $68203$  & 15.61 & 0.49 \\ 
4 XMM [{\bf 8x}] & $0401211001$ &  2006-10-12  &  $47608$  & $67813$  & 27.45 & 0.96 \\ 
\enddata 
\end{deluxetable}

\begin{deluxetable}{rrccccc} 
\rotate
\tabletypesize{\scriptsize}
\tablecolumns{7} 
\tablewidth{0pc} 
\tablecaption{Observations Count Rate and Exposure time of Chandra observations taken in October of 2006 \label{cxo}} 
\tablehead{ 
\colhead{\bf Observation} &  \colhead{ObsID}  & \colhead{Obs. Date} & 
\multicolumn{2}{c}{Exposure time} & \multicolumn{2}{c}{Mean Count Rate} \\
\colhead{} & \colhead{} & \colhead{[Year/Month/Day]} & \multicolumn{2}{c}{[seconds]}  & \multicolumn{2}{c}{[counts/sec] $\pm \sigma$}  \\
\colhead{} & \colhead{} & \colhead{} & \colhead{\bf MEG} & \colhead{\bf HEG} & 
\colhead{\bf MEG} & \colhead{\bf HEG} }
\startdata
1 CXO [{\bf 3c}] & $8452$ &  2006-10-09  &  $19832$  &  $19832$  & 0.62  & 0.32 \\ 
2 CXO [{\bf 4c}] & $7282$ &  2006-10-10  &  $41410$  &  $41410$  & 0.41  & 0.23 \\ 
3 CXO [{\bf 6c}] & $8451$ &  2006-10-11  &  $47360$ &  $47360$  & 0.74  & 0.38 \\ 
4 CXO [{\bf 7c}] & $8450$ &  2006-10-12  &  $38505$  &  $38505$  & 0.79  & 0.39 \\ 
5 CXO [{\bf 9c}] & $7281$ &  2006-10-14  &  $42443$  &  $42443$  & 0.40  & 0.23 \\ 
\enddata 
\end{deluxetable} 

\subsection{XMM-Newton Observatory Data}

We studied all the NGC 3516 XMM-Newton X-ray spectra from the {\it RGS} (Reflection Grating Spectrometer) and the {\it EPIC-PN} (European Photon Imaging Camera). 
The observations were performed between October 6 to October 12 of 2006.

The RGS data were processed with the standard pipeline of Science Analysis System ({\it SAS}) v8.0.1 \citep{gab04}.  We produced source and background spectra, as well 
as response matrices for the RGS data with the {\it rgsproc} task. The RGS spectra were grouped into two channels per bin. We considered the wavelength range from 8 to 
38 \AA{} [0.33 - 1.55 keV], that covers most of the soft X-ray band.

EPIC-PN spectra were also generated using SAS. First, intervals of flaring particle background were selected in order to clean the event list using the method 
presented in \cite{pic04}. The spectra were extracted from a circle with center on the observed position of NGC 3516 and radius $\sim 40''$ in all observations.
The background spectra were obtained from a circular region with similar radius in the same chip of the source. No sign of pile-up was detected in any of the observations as 
reported before by \cite{tur08} and \cite{meh10}. The source and background spectra were generated with the {\em evselect} task. Then, the redistribution matrix and the 
ancillary files were created with the {\em rmfgen} and the {\em arfgen} tasks, respectively. The spectra were grouped to a minimum of 20 counts per bin in order to be able to
use $\chi^{2}$ statistics. We selected the interval from 0.3 keV to 10 keV that includes the soft and hard X-ray bands. In Table ~\ref{xmm} the net count rate and the exposure 
time for each obsID and detector of the XMM-Newton observations of NGC 3516 are shown. 

\subsection{Chandra X-ray Observatory Data}

The Chandra X-ray observations were performed from October 9 to October 14 in 2006 using the High Energy Transmission Grating ({\it HETGS}) with the {\it ACIS-S} 
(Advanced CCD Imaging Spectrometer). The {\it HETGS} \citep{can00} contains two grating assemblies, the Medium Energy Grating ({\it MEG}) and High Energy Grating 
({\it HEG}). We extracted spectra from both gratings using the Chandra Interactive Analysis of Observations software ({\it CIAO v.3.4}, \cite{fru06}) (we followed the 
standard pipeline processes). Negative and positive first-order spectra, and their response matrices were obtained and co-added. The spectra were  grouped into two 
channels per bin.  The count rates and the exposure time in the Medium Energy Grating and High Energy Grating spectra are given in Table~\ref{cxo}.

\section{Data Analysis}

\subsection{Time Variability\label{time_varability}} 

The 2006 X-ray observations of NGC 3516 were performed with almost continuous time coverage (with XMM-Newton and/or Chandra) in a time scale of nine days 
(see Tables ~\ref{xmm} and ~\ref{cxo}). We have numbered the observations in sequential order according to the date of observation, and further included a {\bf x}  
suffix  if performed by XMM-Newton or a {\bf c} suffix if obtained with Chandra. 

There is strong flux variability among different observations with the same observatory (compare count rates for observations with the same observatory/instrument in 
Tables ~\ref{xmm} and ~\ref{cxo}) and within single observations. A detailed light curve of the nine observations can be observed in Figures 1 and 2 of \cite{tur08}. In this 
paper, we study the temporal evolution of the absorbing components in response to these flux changes.

\subsection{XMM-Newton Spectral Analysis} 

We performed the analysis of the XMM-Newton observations using simultaneously both the EPIC-PN and RGS data sets. All spectra were analyzed with the {\it Sherpa} 
\citep{fre01} package included in the CIAO software. Throughout the paper, we attenuate all models with an equivalent H column density N$_{H}$=3.23$\times
$10$^{20}$ cm$^{-2}$ \citep{dic90} to account for the Galactic Absorption in the line of sight towards NGC 3516.    

\subsubsection{Analysis in the X-ray Hard Energy Band}

First, we modelled the NGC 3516  EPIC-PN spectra in the 2.5 - 10 keV band. We fit the data using a redshifted Power Law model plus a Gaussian to account for the 
Fe-K$\alpha$ emission line reported by \cite{tur08} and \cite{meh10}. This model presents strong negative residuals around 7 keV, confirming the two significant 
absorption lines reported by \cite{tur08} and \cite{meh10}, corresponding to Fe K transitions of FeXXV and FeXXVI. These features were modelled with Gaussians 
at this initial stage (although a self consistent photoionization model was applied to this absorption system later on). Table \ref{Table3} lists the best fit parameters for each 
observation. The model is statistically acceptable for all observations (Table \ref{Table3}).

\begin{deluxetable}{rrrrrrrrrr} 
\tabletypesize{\scriptsize}
\tablecolumns{10}
\tablewidth{0pc} 
\tablecaption{Hard band fit: 2.5 - 10 keV energy band. \label{Table3}} 
\tablehead{ 
\colhead{\bf Obs} & \colhead{} & \multicolumn{2}{c}{\bf Power law} & \colhead{} & \colhead{\bf Fe-K$\alpha$} 
& \colhead{\bf FeXXV} & \colhead{\bf FeXXVI} & \colhead{} & \colhead{\bf Statistics} \\
\cline{3-4} \cline{6-8} \\
\colhead{} & \colhead{} & \colhead{ $\Gamma$} & \colhead{ Norm$^{a}$} & \colhead{} 
& \colhead{ Pos[keV]} & \colhead{ Pos[keV]} & \colhead{ Pos[keV]} & \colhead{} 
& \colhead{$\chi_{red}/dof$} \\
\colhead{} & \colhead{} & \colhead{} & \colhead{} & \colhead{} & \colhead{$\sigma$ [keV]} 
& \colhead{ $\sigma$ [keV]} & \colhead{ $\sigma$ [keV]} & \colhead{} & \colhead{\bf [2.5 - 10 keV]} \\
\colhead{} & \colhead{} & \colhead{} & \colhead{} & \colhead{} & \colhead{Norm$^{b}$} 
& \colhead{Norm$^{b}$} & \colhead{Norm$^{b}$} & \colhead{} & \colhead{} }
\startdata
{\bf 1x} &  & $1.64\pm0.02$ & $1.22\pm0.02$ &  & $6.38^{+0.03}_{-0.02}$ & $6.73\pm0.03$ 
& $7.04\pm0.02$  &  & $0.79/1364$ \\
  &  &  &  &  & $0.13^{+0.05}_{-0.03}$ & $0.04\pm0.04$ & $0.05\pm0.04$ &  &  \\   
  &  &  &  &  & $5.89^{+0.64}_{-0.84}$ & $-1.42^{+0.31}_{-0.30}$ & $-1.58\pm0.03$ &  &  \\   
\hline
{\bf 2x} &  & $1.59\pm0.02$ & $0.99\pm0.02$ &  & $6.39\pm0.02$ & $6.703\pm0.019$ & 
$7.05\pm0.02$ &  & $1.08/1411$ \\
  &  &  &  &  & $0.12\pm0.02$ & $0.03\pm0.04$ & $0.05\pm0.04$ &  &  \\   
  &  &  &  &  & $7.37^{+0.43}_{-0.45}$ & $-1.15\pm0.23$ & $-1.38\pm0.23$ &  &  \\   
\hline
{\bf 5x} &  & $1.45\pm0.02$ & $0.65\pm0.02$ &  & $6.38\pm0.02$ & $6.75\pm0.02$ &
$7.08\pm0.02$ &  & $1.17/1357$ \\ 
  &  &  &  &  & $0.13\pm0.02$ & $0.02\pm0.04$ & $0.03\pm0.03$ &  &  \\   
  &  &  &  &  & $7.29^{+0.44}_{-0.45}$ & $-1.06\pm0.21$ & $-1.16\pm0.22$ &  &  \\ 
\hline
{\bf 8x} &  & $1.54\pm0.02$ & $0.95\pm0.02$ &  & $6.41\pm0.02$ & $6.78\pm0.02$ & 
$7.11^{+0.03}_{-0.02}$ &   & $0.79/1483$ \\ 
  &  &  &  &  & $0.09\pm0.02$ & $0.04\pm0.03$ & $0.05\pm0.03$  &  \\   
  &  &  &  &  & $5.89^{+0.48}_{-0.47}$ & $-1.84\pm0.31$ & $-1.86\pm0.32$  &  \\ 
\enddata 
\tablenotetext{a}{ in $10^{-2}$ $photons/keV/{cm^{2}}/s$ at 1 keV}
\tablenotetext{b}{ in $10^{-5}$ $photons/{cm^{2}}/s$ in the line}
\end{deluxetable} 

\subsubsection{X-ray Broad Band Analysis}

We fit the EPIC-PN and RGS simultaneously,  extrapolating the fit in the hard band to the entire spectrum [0.3 - 10 keV]. A soft X-ray excess is evident in the residuals. We fit 
this emission feature with a Black Body. Initially, we left the temperature of the Black Body free to vary independently among observations. However, since in all XMM-
Newton observations the kT value is always around 0.1 keV (see Table \ref{Table4}), this parameter was free to vary but constrained to have the same value in all 
observations (the temperature of the Black Body was linked to a single best fit value among the all observations). The normalization was fit independently in each 
observation. 

This model presents strong residuals in the soft band typical of ionized  absorption, due to the well know warm absorber in this source. 

\begin{deluxetable}{rrrrrrrrrrr} 
\tabletypesize{\scriptsize}
\tablecolumns{7} 
\tablewidth{0pc} 
\tablecaption{Whole band fit: 0.3 - 10 keV.\label{Table4}} 
\tablehead{ 
\colhead{\bf Obs} & \colhead{} & \colhead{\bf Power law} & \colhead{} & \colhead{\bf Black Body} & \colhead{} 
& \colhead{\bf Statistics} \\
\colhead{} & \colhead{} & \colhead{$\Gamma$} & \colhead{} & \colhead{kT [keV]} & \colhead{} 
& \colhead{$\chi_{red}$/dof} \\
\colhead{} & \colhead{} & \colhead{Norm$^{a}$} & \colhead{} & \colhead{Norm$^{c}$} &  \colhead{} 
& \colhead{\bf [0.3,10 keV]} }
\startdata
{\bf 1x} &  & $1.31\pm0.02$ &  & $0.08\pm0.02$ &  & $4.5/4534$ \\ 
  &  & $0.78\pm0.02$ &  & $5.92\pm0.05$ &  &  \\  
\hline
{\bf 2x} &  & $1.30\pm0.02$ &  & $0.09\pm0.02$ &  & $5.2/4581$ \\ 
  &  & $0.66\pm0.02$ &  & $5.85\pm0.04$ &  &  \\  
\hline
{\bf 5x} &  & $1.05\pm0.02$ &  & $0.09\pm0.02$ &  & $4.2/4527$ \\ 
  &  & $0.36\pm0.02$ &  & $2.99\pm0.03$ &  &  \\  
\hline
{\bf 8x} &  & $1.21\pm0.02$ &  & $0.09\pm0.02$ &  & $5.2/4655$ \\ 
  &  & $0.61\pm0.02$ &  & $6.38\pm0.04$ &  &  &  \\  
\enddata 
\tablenotetext{a}{ in $10^{-2}$ $photons/keV/{cm^{2}}/s$ in al 1 keV}
\tablenotetext{c}{ in $10^{-4}$ $L_{39}/D_{10}^{2}$ units, $L_{39}$ is the source luminosity in units of 
$10^{39}$ $ergs/sec$ and $D_{10}$ is the distance to the source in units of 10 kpc}
\end{deluxetable}

\subsubsection{Modelling the Ionized Absorber in NGC 3516}

There have been several studies of the warm absorbers in NGC 3516, reporting a different number of ionization components. In order to establish the presence of each 
component, we decided to add one absorber component at a time in our models. This allows us to test statistically its presence, and also to inspect visually its contribution to 
the overall opacity.

\begin{figure}
   \centering
   \includegraphics[width=1.25\linewidth]{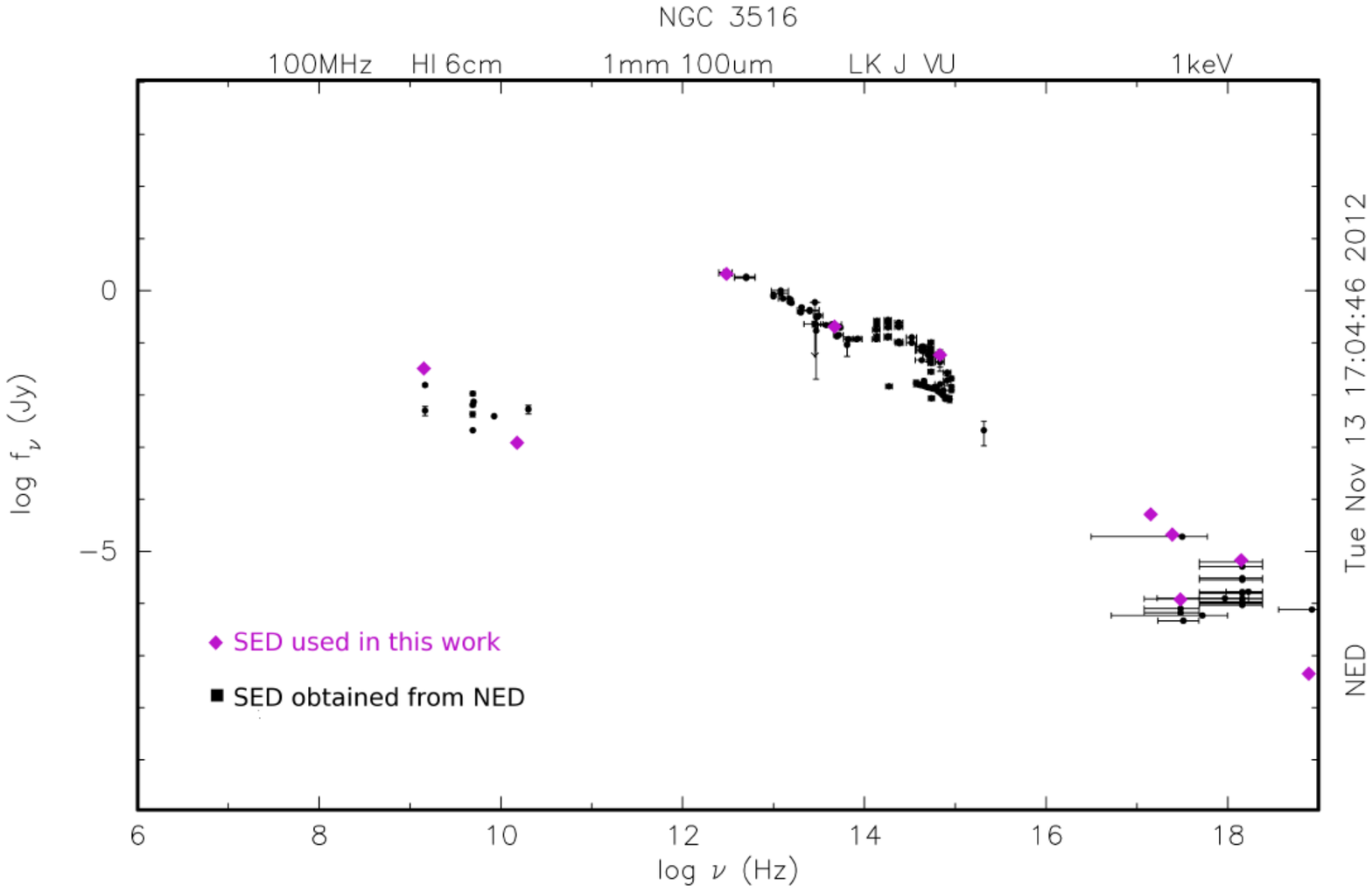}
   \caption{Spectral Energy Distribution of NGC 3516. The black dots are reported by NED while in this work, we derived the fuchsia  
points values. \label{fig1}}
\end{figure}

The {\it PHASE} code \citep{kro03} was used to fit the warm absorbers (WA). The model has four free parameters: 1) The ionization parameter U (U = $\frac {Q}{4{\pi}
R^{2}n_{H}c}$) \footnote{Ionization parameter: U = $\frac {Q}{4{\pi}R^{2}n_{H}c}$, where Q is the rate of 0.013-100 keV photons produced in the source, R the distance 
between the black hole and the accretion disk system to WA location, n$_{H}$ the Hydrogen number density and c the light speed.} ; 2) the  equivalent Hydrogen column 
density N$_{H}$; 3) the outflow velocity v$_{z}$; and 4) the internal micro-turbulence velocity v$_{turb}$. The spectral energy distribution ({\it SED}) of the source, also 
required in the models, was obtained from {\it NED} (NASA/IPAC Extragalactic Database) between the radio and the UV regimes, and complemented with our fits to the 
X-ray band. The SED is shown in Figure \ref{fig1}. 

We note that the estimated errors in the outflow velocity of the absorber are too small, in general, smaller than the resolution of the detectors. For this reason, throughout the 
paper we consider a conservative error in this parameter as half the minimum spectral resolution of the RGS (700 kms$^{-1}$ in 15 \AA): ${\Delta}v_{out}$ = 350 km s$^{-1}$. 
We prefer this conservative value given that the fits to the data include several blended absorption lines from different charge states as well as the blend of several velocity
components present in absorption.

Our first model ({\bf Model A}) consists of a single ionization component, and an F-test confirms its presence  with $>99.99\%$ confidence level for all observations. This 
component models absorbing lines produced mainly by OVII, OVIII, as well as part of the  Fe M-shell unresolved transition array ({\it UTA}),  among others. The outflow velocity 
is  -532$\pm$350 kms$^{-1}$.

{\bf Model B} included a second WA absorbing phase. A statistically better fit than {\bf Model A} is supported by an F-test ($>99.99\%$ of significance for all XMM-Newton
 observations). The second absorber fits absorption lines with higher level of ionization, mainly Fe L-shell absorption from charge states XVII to XXII, also including 
absorption by NeX. The outflow velocity is higher also, -1845$\pm$350 kms$^{-1}$. We note that the two ionized absorption phases had different U and N$_{H}$ values 
among observations. We will analyze these changes in section \ref{variability}.  
 
Given that residuals were still present, we further included a third WA phase ({\bf Model C}). At the beginning, all the parameters were free to vary. However, for each 
absorbing component, the best-fit value of the outflow velocity v$_{out}$ among the observations was similar, with differences lower than the RGS resolution: 700 kms$^{-1}$ 
(in 15 \AA). Thus, we left the outflow velocity of each absorber as a free parameter, but constrained it to have the same best-fit value in all the observations (we 
link this parameter in the models for all observations, assuming no acceleration or deceleration in the flow during the total observing time). Given that the best fit values of the 
turbulent velocity v$_{turb}$ were also very similar among observations, with differences lower than RGS resolution, we performed the same procedure with the turbulent 
velocity, linking this free parameter to a single value in all datasets. According to an F-test, the third absorber included in model C is statistically required by the data 
(probability $>99.99\%$). This third absorber is modelling low ionization lines produced by charge states such as NeV, and NeVI, and a fraction of the Fe M-shell UTA. This 
component had an outflow velocity best-fit value higher than the other two absorbers, v$_{out}=$-2425$\pm$350 kms$^{-1}$.

Thus, this final model includes of 3 absorbing components. The first one consists of a medium level of  ionization (hereafter phase {\bf MI}); the 
second phase presents high ionization (hereafter phase {\bf HI}) and the third one is produced by low ionized gas (hereafter phase {\bf LI}). 
Tables \ref{Table5} and \ref{Table6} summarize the models applied in the four XMM-Newton spectra.

Although the data does not present strong residuals consistent with additional absorption features, we tested a possible fourth WA component in the soft X-ray RGS data. 
An F-test does not show a statistic improvement over Model C.   

\begin{deluxetable}{rrrrrrrrr} 
\rotate
\tabletypesize{\scriptsize}
\tablecolumns{9} 
\tablewidth{0pc} 
\tablecaption{Models to fit NGC 3516 in XMM-Newton observations\label{Table5}} 
\tablehead{ 
\colhead{\bf Model} & \multicolumn{2}{c}{\bf Power Law} &  \multicolumn{2}{c}{\bf Black Body} 
& \multicolumn{3}{c}{\bf WA parameters} & \colhead{\bf Statistics} \\
\cline{2-8} \\
\colhead{} & \colhead{$\Gamma$} & \colhead{Norm$^{a}$} & \colhead{KT} 
& \colhead{Norm$^{c}$} & \colhead{Phase {\bf LI}} & \colhead{Phase {\bf MI}} & \colhead{Phase {\bf HI}} 
& \colhead{$\chi_{red}$/dof}     } 
\startdata
{Obs {\bf 1x}} &  &  &  &  &  &  &  & \\
\hline
{\bf A} & $1.69\pm0.02$ & $1.36\pm0.02$ & $0.10\pm0.02$ & $9.18\pm0.06$ &  & $\surd$ & & 
$1.44/4530$ \\
{\bf B} & $1.78\pm0.02$ & $1.58\pm0.04$ & $0.10\pm0.02$ & $5.29^{+0.04}_{-0.09}$ &  & $\surd$ & 
$\surd$ & $0.95/4526$ \\
{\bf C} & $1.81\pm0.02$ & $1.67\pm0.06$ & $0.09\pm0.02$ & $6.47\pm0.02$ & $\surd$ & $\surd$ & 
$\surd$ & $0.86/4504$ \\
\hline
{Obs {\bf 2x}} &  &  &  &  &  &  &  & \\
\hline
{\bf A} & $1.66\pm0.02$ & $1.11\pm0.02$ & $0.10\pm0.02$ & $5.82^{+0.08}_{-0.05}$ & 
& $\surd$ &  & $1.73/4577$ \\
{\bf B} & $1.71\pm0.02$ & $1.23\pm0.02$ & $0.10\pm0.02$ & $5.78\pm0.04$ &  & $\surd$ & 
$\surd$ & $1.16/4573$ \\
{\bf C} & $1.74\pm0.05$ & $1.31\pm0.09$ & $0.09\pm0.02$ & $6.50\pm0.40$ & $\surd$ & $\surd$ & $\surd$ & 
$1.04/4557$ \\
\hline
{Obs {\bf 5x}} &  &  &  &  &  &  &  & \\
\hline
{\bf A} & $1.47\pm0.02$ & $0.68\pm0.02$ & $0.10\pm0.02$ & $5.82^{+0.08}_{-0.05}$ & & $\surd$ &  
& $1.23/4523$  \\
{\bf B} & $1.54\pm0.02$ & $0.79\pm0.02$ & $0.10\pm0.02$ & $3.06^{+0.03}_{-0.05}$ &  & 
$\surd$ & $\surd$ & $0.98/4519$ \\
{\bf C} & $1.61\pm0.02$ & $0.89\pm0.02$ & $0.09\pm0.02$  & $4.39\pm0.04$ & $\surd$ & 
$\surd$ & $\surd$ & $0.88/4503$ \\
\hline 
{Obs {\bf 8x}} &  &  &  &  &  &  &  & \\
\hline
{\bf A} & $1.63\pm0.02$ & $1.13\pm0.02$ & $0.10\pm0.02$ & $8.09\pm0.05$ &  & $\surd$ &  & 
$1.74/4649$ \\
{\bf B} & $1.70\pm0.02$ & $1.28\pm0.02$ & $0.09\pm0.02$ & $6.23^{+0.05}_{-0.03}$ &  & $\surd$ & 
$\surd$ & $1.12/4645$ \\
{\bf C} & $1.75\pm0.05$ & $1.4\pm0.2$ & $0.09\pm0.02$ & $7.9\pm0.4$ & $\surd$ & $\surd$ & $\surd$ & 
$0.97/4629$ \\
\enddata 
\tablenotetext{a}{In $10^{-2}$ $photons/keV/{cm^{2}}/s$ in at 1 keV}
\tablenotetext{b}{In $10^{-4}$ $L_{39}/D_{10}^{2}$ units, $L_{39}$ is the source luminosity in units of 
$10^{39}$ $ergs/sec$ and $D_{10}$ is the distance to the source in units of 10 kpc} 
\end{deluxetable}

\begin{deluxetable}{rrrrrrrrrr} 
\rotate
\tabletypesize{\scriptsize}
\tablecolumns{10} 
\tablewidth{0pc} 
\tablecaption{WA parameters of NGC 3516. \label{Table6}} 
\tablehead{ 
\colhead{\bf Model C} & \colhead{} & \multicolumn{2}{c}{Phase {\bf LI}} & \colhead{} & \multicolumn{2}{c}{Phase {\bf MI}} & \colhead{} 
&\multicolumn{2}{c}{Phase {\bf HI}} \\
\cline{3-4} \cline{6-7} \cline{9-10} \\
\colhead{} & \colhead{} &\colhead{logU} & \colhead{logN$_{H}$} & \colhead{} & \colhead{logU} 
& \colhead{log$N_{H}$} & \colhead{} & \colhead{logU} & \colhead{logN$_{H}$} \\ 
\colhead{} & \colhead{} &\colhead{vel$_{z}$ [kms$^{-1}$]} & \colhead{vel$_{turb}$ [kms$^{-1}$]} & \colhead{} 
& \colhead{vel$_{z}$ [kms$^{-1}$]} & \colhead{vel$_{turb}$ [kms$^{-1}$]} & \colhead{} & \colhead{vel$_{z}$ [kms$^{-1}$]} 
& \colhead{vel$_{turb}$ [kms$^{-1}$]} }
\startdata
{Obs {\bf 1x}} &  &  &  &  &  &  &  &  & \\
\hline
{\bf A} &  &  &  &  & $0.29^{+0.03}_{-0.02}$ & $21.82 \pm 0.02$ &  &  &  \\
 &  &  &  &  & $-888 \pm 350$ & $286 \pm 350$ &  &  &  \\
\hline
{\bf B} &  &  &  &  & $0.16 \pm 0.02$ & $21.65 \pm 0.02$ &  & $1.65 ^{+0.12}_{-0.02}$ & 
$22.07 \pm 0.02$ \\
 &  &  &  &  & $-995 \pm 350$ & $220  \pm 350$ &  & $-2302 \pm 350$ & $320 \pm 350$   \\
\hline
{\bf C} &  & $-0.93 ^{+0.03}_{-0.02}$ & $21.14 \pm 0.02$ &  & $0.38 \pm 0.02$ & $21.56 \pm 0.02$ &  
& $1.69 \pm 0.02$ & $22.25 \pm 0.02$ \\  
  &  & $-2425 \pm 350$ & $88 \pm 350$ &  & $-532 \pm 350$ & $239\pm350$ &  & $-1845 \pm 350$ & 
  $112 \pm 350$ \\ 
\hline
\hline
{Obs {\bf 2x}} &  &  &  &  &  &  &  &  &  \\
\hline
{\bf A} &  &  &  &  & $0.27 \pm 0.02$ & $21.82 \pm 0.02$ &  &  &  \\ 
 &  &  &  &  & $-1018 \pm 350$ & $207 \pm 350$   &  &  &  \\ 
\hline
{\bf B} &  &  &  &  & $0.18 \pm 0.02$ & $21.68 \pm 0.02$ &  & $1.95 \pm 0.02$ & $22.29 \pm 0.08$ \\
 &  &  &  &  & $-1082 \pm 350$ & $195 \pm 350$ &  & $-1851 \pm 350$ & $77 \pm 350$ \\
\hline
{\bf C} &  & $-0.78 \pm 0.05$ & $20.98 \pm 0.08$ & & $0.27 \pm 0.02$ & $21.52 \pm 0.09$ &  & 
$1.68 \pm 0.09$ & $22.197 \pm 0.164$ \\
 &  & $=>$ {\bf 1x} & $=>$ {\bf 1x} &  & $=>$ {\bf 1x} & $=>$ {\bf 1x} &  & $=>$ {\bf 1x} & $=>$ 
 {\bf 1x} \\
\hline
\hline
Obs {\bf 5x} &  &  &  &  &  &  &  &  &  \\
\hline
{\bf A} &  &  &  &  & $0.215\pm0.024$ & $21.94\pm0.02$ &  &  &  \\   
 &  &  &  &  & $-1388\pm 350$ & $168\pm350$ &  &  &  \\   
\hline
{\bf B} &  &  &  &  & $0.12\pm0.02$ & $21.75\pm0.02$ &  & $1.68\pm0.02$ & $22.22\pm0.03$ \\
 &  &  &  &  & $-1066\pm 350$ & $199\pm350$ &  & $-885\pm 350$ & $143\pm350$ \\ 
\hline
{\bf C} &  & $-1.22\pm0.02$ & $21.25\pm0.02$ &  & $0.29\pm0.02$ & $21.62\pm0.02$ &  & 
$1.77\pm0.02$ & $22.33\pm0.14$ \\
 &  & $=>$ {\bf 1x} & $=>$ {\bf 1x} &  & $=>$ {\bf 1x} & $=>$ {\bf 1x} &  & $=>$ {\bf 1x} & $=>$ 
 {\bf 1x} \\
\hline
\hline
{Obs {\bf 8x}} &  &  &  &  &  &  &  &  &  \\
\hline
{\bf A} &  &  &  &  & $0.33\pm0.02$ & $21.93\pm0.02$ &  &  &  \\  
 &  &  &  &  & $-1154 \pm 350$ & $202\pm350$  &  &  &  \\ 
\hline
{\bf B} &  &  &  &  & $0.18\pm0.03$ & $21.69\pm0.02$ &  & $1.77\pm0.02$ & $22.07\pm0.02$ \\
 &  &  &  &  & $-619\pm 350$ & $234.9\pm350$ &  & $-2293\pm 350$ & $417 \pm 350$ \\
\hline
{\bf C} &  & $-0.78\pm0.02$ & $21.25\pm0.09$ &  & $0.38\pm0.02$ & $21.512\pm0.032$ &  & $1.96\pm0.22$ & 
$22.48\pm0.22$ \\
 &  & $=>$ {\bf 1x} & $=>$ {\bf 1x} &  & $=>$ {\bf 1x} & $=>$ {\bf 1x} &  & $=>$ {\bf 1x} & $=>$ 
 {\bf 1x} \\  
\enddata 
\tablenotetext{a}{in $10^{-2}$ $photons/keV/{cm^{2}}/s$ in at 1 keV}
\tablenotetext{b}{in $10^{-4}$ $L_{39}/D_{10}^{2}$} 
\tablenotetext{*}{In model C we referred the v$_{out}$ and v$_{out}$ to the values of observation {\bf 1x}, however the final values
were fitted taking into account the four XMM-Newton observations.}
\end{deluxetable} 

\begin{figure}
  \centering
 \includegraphics[width=.6\textwidth]{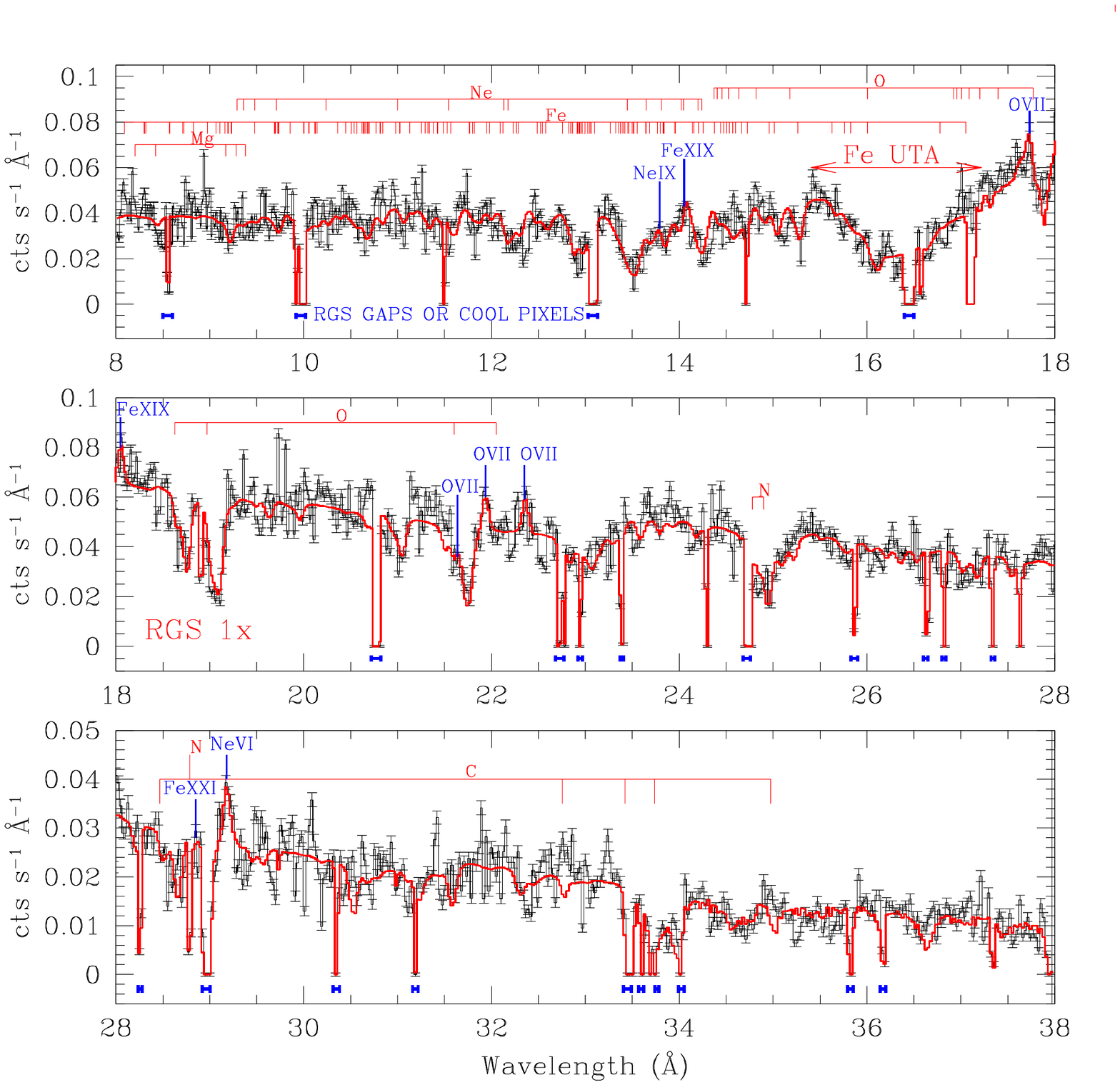}
\caption{RGS spectrum of observation {\bf 1x} together with the {\bf Model C} fit -red- 
[8 - 38 \AA{}]. The absorption lines are identified above with the corresponding transition and the 
main emission lines are marked in blue (top on the panel). The detector gaps the cool pixels of RGS
are marked in blue on the bottom of the panel (http://heasarc.gsfc.nasa.gov/docs/xmm/uhb
rgsmultipoint.html).\label{fig2}}
\end{figure}

\begin{figure}
 \centering
 \includegraphics[width=.6\textwidth]{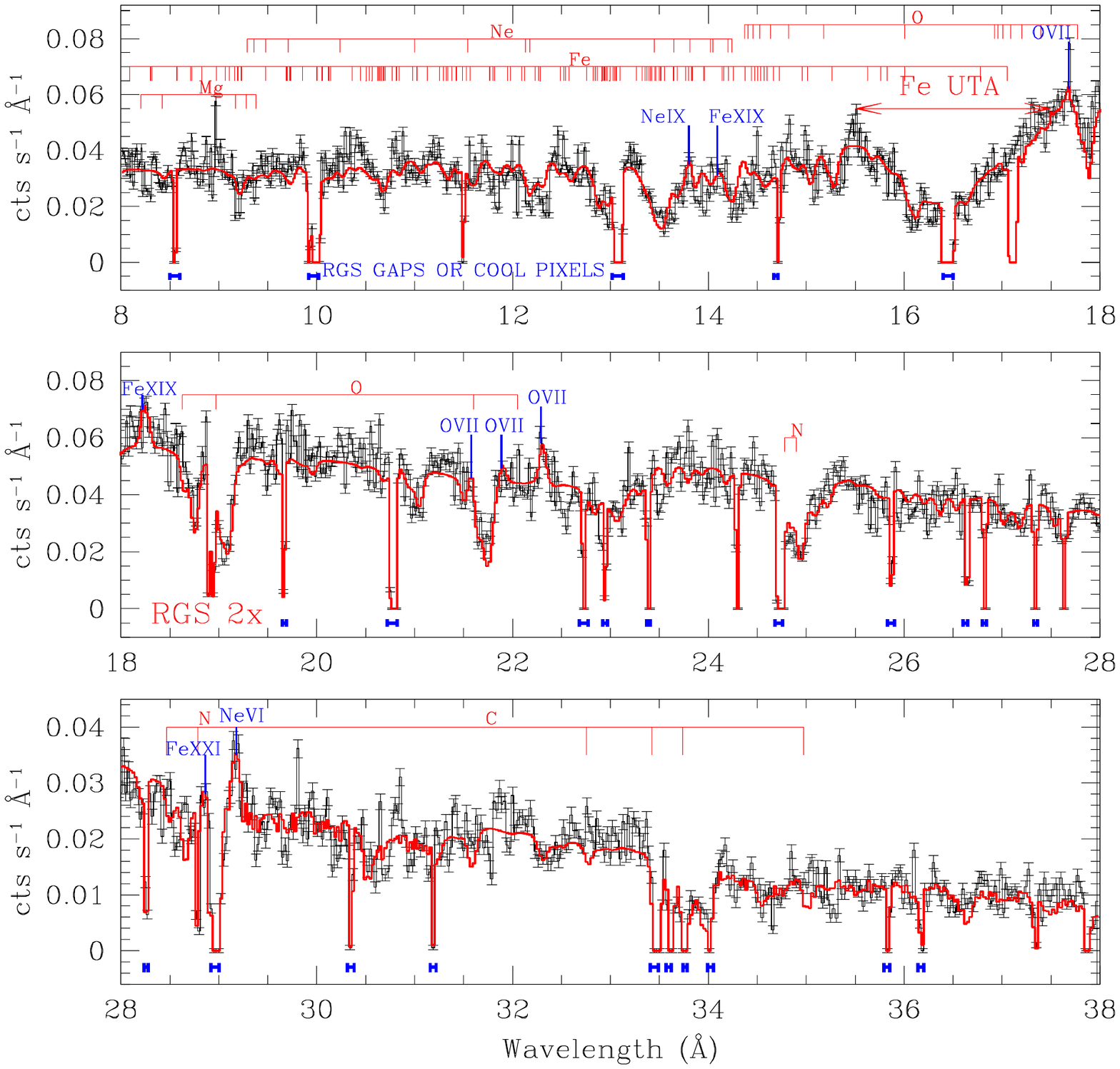}
\caption{RGS spectrum of observation {\bf 2x} together with the {\bf Model C} fit -red- 
[8 - 38 \AA{}]. The absorption lines are identified above, the corresponding transition and the main 
emission lines are marked in blue (top on the panel). The detector gaps or the cool pixels of RGS
are marked in blue on the bottom of the panel (http://heasarc.gsfc.nasa.gov/docs/xmm/uhb
rgsmultipoint.html).\label{fig3}}
\end{figure}

\begin{figure}
\centering
 \includegraphics[width=.6\textwidth]{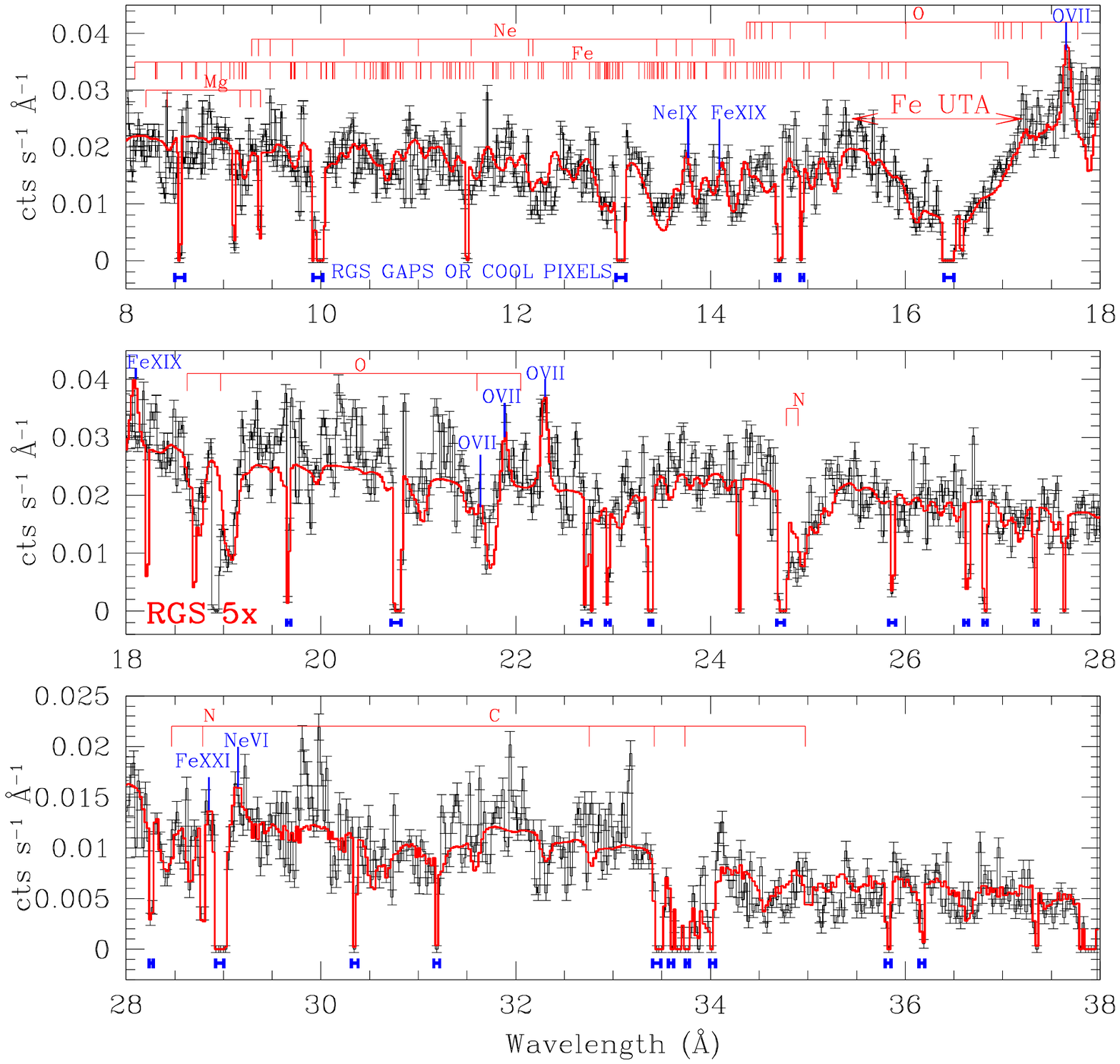}
\caption{RGS spectrum of observation {\bf 5x} together with the {\bf Model C} fit -red- 
[8 - 38 \AA{}]. The absorption lines are identified above, the corresponding transition and the main 
emission lines are marked in blue (top on the panel). The detector gaps or the cool pixels of RGS
are marked in blue on the bottom of the panel (http://heasarc.gsfc.nasa.gov/docs/xmm/uhb
rgsmultipoint.html).\label{fig4}}
\end{figure}

\begin{figure}
 \centering
 \includegraphics[width=.57\textwidth]{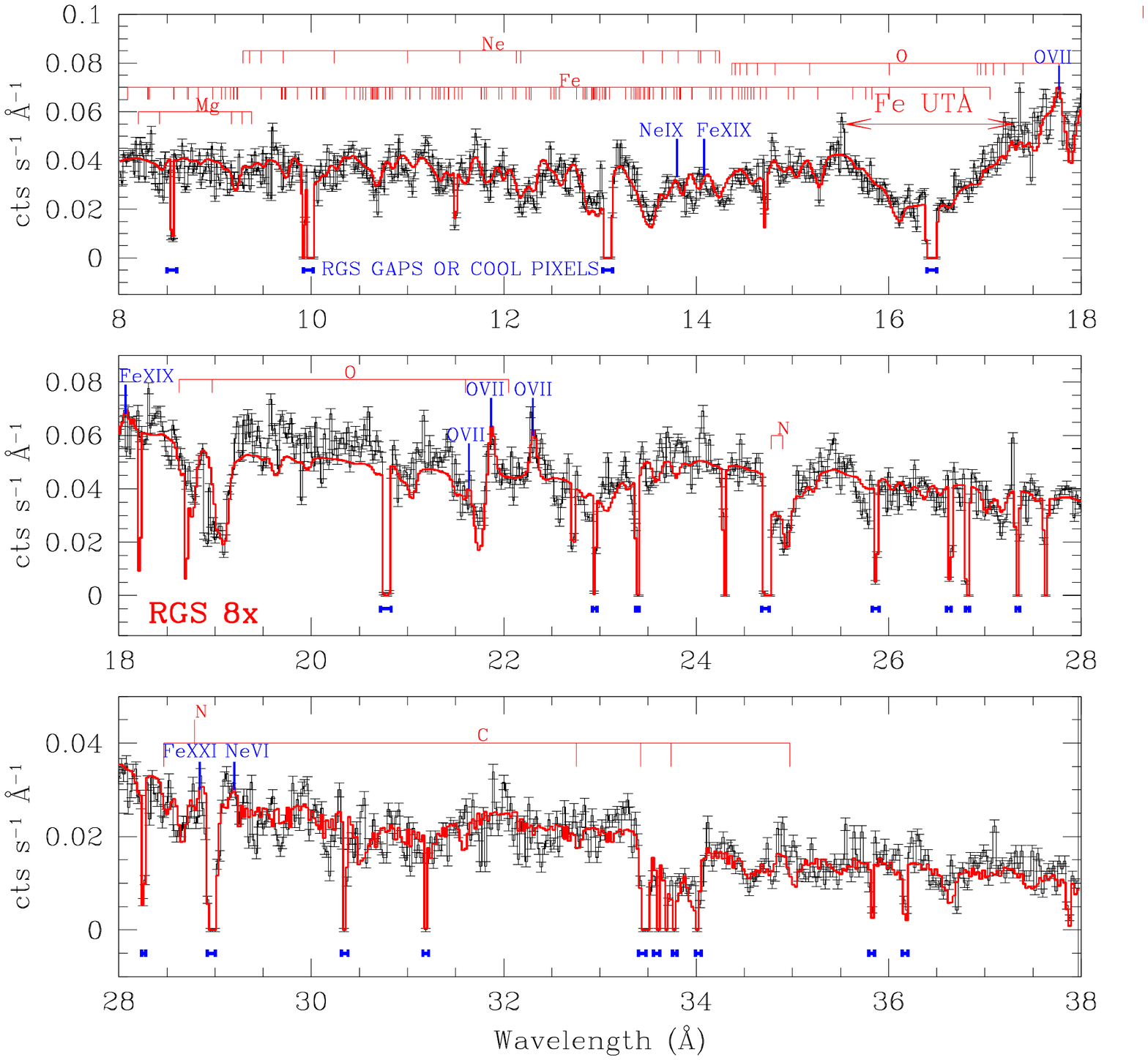}
\caption{RGS spectrum of observation {\bf 8x} together with the {\bf Model C} fit -red- 
[8,38 \AA{}]. The absorption lines are identified above, the corresponding transition and the main 
emission lines are marked in blue (top on the panel). The detector gaps or the cool pixels of RGS
are marked in blue on the bottom of the panel (http://heasarc.gsfc.nasa.gov/docs/xmm/uhb
rgsmultipoint.html).\label{fig5}}
\end{figure}

\begin{deluxetable}{rrrrrrrrrr} 
\rotate
\tabletypesize{\scriptsize}
\tablecolumns{10} 
\tablewidth{0pc} 
\tablecaption{Emission Lines in the Soft Band of XMM-Newton and Chandra. \label{Table7}} 
\tablehead{ 
\colhead{\bf Transition$^{a}$} &  \colhead{\bf NeIX} & \colhead{\bf FeXIX} & \colhead{\bf OVII} 
&  \colhead{\bf FeXIX} & \colhead{\bf OVII} & \colhead{\bf OVII} & \colhead{\bf OVII} & \colhead{\bf FeXXI} & \colhead{\bf NeVI}  \\ 
\colhead{} & \colhead{$\lambda13.69^{*}$} & \colhead{$\lambda13.91^{*}$} & \colhead{$\lambda17.39^{*}$} 
&  \colhead{$\lambda17.86^{*}$} & \colhead{$\lambda21.60^{*}$} & \colhead{$\lambda21.81^{*}$} & 
\colhead{$\lambda22.10^{*}$} & \colhead{$\lambda28.53^{*}$} & \colhead{$\lambda28.92^{*}$}  \\ 
\cline{2-10}\\ 
 \colhead{} & \multicolumn{9}{c}{Wavelength Observed [{\bf \AA}]} \\
\colhead{\bf Observation} & \multicolumn{9}{c}{Energy Flux$^{b}$}  }
\startdata
{\bf 1x} & $13.79\pm0.08$ & $14.04\pm0.04$ & $17.73\pm0.02$	& $18.05\pm0.12$ & $21.64\pm0.02$ & $21.93\pm0.02$ & $22.29\pm0.22$ & $28.83\pm0.03$ & $29.19\pm0.02$ \\ 
	     & $1.4\pm0.5$ & $2.4\pm0.6$ & $1.9\pm0.05$ & $1.9\pm0.05$ & $9.8\pm3.2$ & $3.9\pm...$ & $0.9\pm0.8$ & $0.5\pm0.2$ & $2.6\pm0.7$ \\ 
{\bf 2x} & $13.79\pm0.02$ & $14.04\pm0.02$ & $17.67\pm0.03$ & $18.19\pm0.03$ & $21.59\pm0.02$ & $21.89\pm0.08$ & $22.29\pm0.09$ & $28.86\pm0.02$ & $29.18\pm0.02$ \\ 
	     & $2.4\pm0.5$ & $1.6\pm0.5$ & $0.9\pm0.4$ & $1.2\pm0.5$ & $0.9\pm0.8$ & $1.2\pm0.9$ & $2.9\pm0.7$ & $0.2\pm0.5$ & $2.9\pm0.7$ \\ 
{\bf 3c} & $13.81\pm0.10$ & $14.06\pm0.02$ & $17.67\pm...$ & $18.16\pm0.72$ & $21.6\pm...$ & $21.91\pm...$ & $22.30\pm...$ & O.R.$^{c}$ & O.R.  \\ 
            & $1.5\pm0.9$ & $2.6\pm1.3$ & $1.5\pm1.8$ & $2.2\pm2.1$ & $0\pm...$ & $3.4\pm...$ & $3.4\pm...$ &  &  \\ 
{\bf 4c} & $13.82\pm0.04$ & $14.06\pm0.06$ & $17.74\pm0.08$ & $18.215\pm$ & $21.59\pm...$ & $21.82\pm$ & $22.31\pm...$ & O.R. & O.R. \\
            & $0.8\pm0.5$ & $1.07\pm0.7$ & $0.7\pm0.6$ & $0.7\pm0.9$ & $1.5\pm...$ & $7.7\pm7.2$ & $6.2\pm5.5$ &  &  \\
{\bf 5x} & $13.79\pm0.02$ & $14.04\pm0.02$ & $17.65\pm0.02$ & $18.16\pm...$ & $21.64\pm$ & $21.89\pm0.2$ & $22.29\pm0.02$ & $28.82\pm...$ & $29.19\pm0.04$ \\
            & $0.9\pm0.3$ & $2.7\pm1.2$ & $1.2\pm0.4$ & $0.7\pm0.5$ & $8.2\pm3.3$ & $1.6\pm0.4$ & $1.9\pm0.5$ & $0.4\pm0.6$ & $1.2\pm0.5$ \\
{\bf 6c} & $13.82\pm0.04$ & $14.06\pm0.02$ & $17.64\pm...$ & $18.11\pm0.10$ & $21.62\pm0.02$ & $21.90\pm0.04$ & $22.31\pm...$ & O.R. & O.R. \\ 
            & $0.9\pm0.6$ & $2.2\pm0.8$ & $0.4\pm0.4$ & $2.6\pm1.5$ & $709\pm645$ & $3.6\pm2.2$ & $4.5\pm2.6$ &  &  \\ 
{\bf 7c} & $13.81\pm...$ & $14.05\pm0.02$ & $17.71\pm0.04$ & $18.08\pm...$ 	& N.D.$^{e}$	& $21.92\pm0.36$ & $22.26\pm$ & O.R. & O.R. \\ 
            & $0.6\pm0.7$ & $1.4\pm0.9$ & $2.2\pm1.4$ & $1.02\pm1.33$ & $0\pm...$ 	& $2.2\pm3.2$ & $1.2\pm6.2$ &  &  \\ 
{\bf 8x} & $13.79\pm...$ & $14.04\pm$ & $17.78\pm0.02$ & $18.07\pm0.15$ & $21.62\pm0.12$ & $21.88\pm$ & $22.31\pm0.02$ & N.D. & $29.21\pm.08$ \\ 
	     & $1.1\pm1.9$ & $0.8\pm0.9$ & $0.6\pm0.5$ & $6.2\pm4.8$ & $123\pm115$ & $2.4\pm0.6$ & $1.99\pm0.75$ 	& $0\pm...$ & $2.2\pm1.7$ \\ 
{\bf 9c} & $13.82\pm0.2$ & $14.07\pm1.8$ & $17.714\pm0.12$ & $18.02\pm0.7$ & N.D.	& N.D. & $22.3\pm...$ & O.R. & O.R. \\ 
	     & $0.9\pm1.3$ & $3.6\pm4.2$ & $1.15\pm1.3$ & $1.6\pm1.5$	& $0\pm...$ & $0\pm...$ & $1.6\pm...$  &   &  \\ 
\enddata 
\tablenotetext{a}{Ion Name and Transition Rest-Frame Wavelength (\AA{})$^{*}$}
\tablenotetext{b}{in $10^{-4}$ $photons/{cm^{2}}/s$ in the line}
\tablenotetext{c}{{\bf ...} means indeterminate.}
\tablenotetext{d}{{\bf O.R.} means line out of detector range.}
\tablenotetext{e}{{\bf N.D.} means line not detected.}
\end{deluxetable} 

\subsubsection{Emission Features in the Spectra} 

The presence of positive residuals in the spectra, associated to emission lines in the rest frame of the object, is evident in the data. Through a detailed visual inspection we 
identified nine emission features using  ATOMDB version 1.3.1 from the Chandra X-ray Center (http://asc.harvard.edu/atomdb/WebGUIDE/), and modelled them with 
Gaussian components. The Full Width at Half Maximum ({\it FWHM}) was fixed to 300 kms$^{-1}$, given that they are not resolved. We found two emission lines 
corresponding with Ne transitions (NeVI and NeIX), five corresponding to OVII, three of them belong to the triplet between 21.6-22.1 \AA{}, and two lines of highly ionized Iron 
(FeXIX and FeXXI). Table \ref{Table7} shows the nine emission lines detected in the soft X-ray band for each observation.  

We stress that not all the emission lines were detected in a significant way in all the observations. However, the fluxes of all lines are consistent with each other among all 
the observations, with the only exception of the OVII-K$\alpha$ transition at $\lambda$ 21.6 \AA. The differences in this transition are probably due to the strong blending 
of this line with the strong absorption feature produced by the WA. We point out that \cite{meh10} only modelled one emission line in the soft X-ray energy band, namely 
OVII (f). These authors only include this feature as it is the only one significant in all observations. Nevertheless, in  their Figure 6 it is possible to observe residuals 
coincident with the emission lines identified here. For instance at wavelengths around $\lambda$ $\sim$13.8,  $\sim$17.8,  $\sim$18, $\sim$22.2 and $\sim$ 29.2 \AA{}.

Summarizing, our final model consists of  {\bf Model C} (including the 3 absorbing component in the soft band) plus the fourth absorbing phase ({\bf VH}) in the hard band, 
plus the nine emission lines described above. The figures with the final models (including a simultaneous fit to the Chandra and XMM data, see \S \ref{simul}) are presented 
in Figures \ref{fig2}, \ref{fig3}, \ref{fig4}, \ref{fig5} and  \ref{fig7}. Table \ref{flux} contains the observed energy flux values for all observations. The flux was integrated in two 
different energy ranges: in the soft energy band [8 - 25 \AA{} : 0.49 - 1.55 keV] and in the hard energy band [1.55 - 10 keV]. The  reported quantities correspond to the intrinsic
flux of the source, without attenuation by Galactic absorption (see \S 3.2) and the warm absorber components. 

\begin{deluxetable}{rrcrc} 
\tabletypesize{\scriptsize}
\tablecolumns{5} 
\tablewidth{0pc} 
\tablecaption{Energy flux [$F_{h\nu}$] of NGC 3516: soft X-ray band and hard X-ray band \label{flux}.} 
\tablehead{ 
\colhead{\bf Obs} &  & \colhead{\bf Soft band} &  & \colhead{\bf Hard band} \\
\colhead{} &  & \colhead{$F_{h\nu}^{*}$ {\bf [8 - 25 \AA : 0.49 - 1.55 keV]}} &  & \colhead{$F_{h\nu}^{*}$ 
{\bf [1.55 - 10 keV]}} }
\startdata
{\bf 1x} &  & $1.87\pm0.05$ &  & $2.59\pm0.03$ \\ 
{\bf 2x} &  & $1.69\pm0.03$ &  & $2.34\pm0.02$ \\ 
{\bf 3c} &  & $1.09\pm0.04$ &  & $2.04\pm0.03$ \\ 
{\bf 4c} &  & $0.707\pm0.019$ &  &$1.58\pm0.03$ \\ 
{\bf 5x} &  & $0.88\pm0.03$ &  & $1.97\pm0.02$ \\ 
{\bf 6c} &  & $1.54\pm0.03$ &  & $2.22\pm0.05$ \\ 
{\bf 7c} &  & $1.72\pm0.03$ &  & $2.13^{+0.08}_{-0.06}$ \\ 
{\bf 8x} &  & $1.84\pm0.04$ &  & $2.46\pm0.03$ \\ 
{\bf 9c} &  & $0.50\pm0.02$ &  & $0.35^{+0.03}_{-0.02}$ \\ 
\hline
\enddata 
\tablenotetext{*}{In $10^{-11}$ $ergs/{cm^{2}}/s$.}
\end{deluxetable}

\subsection{Simultaneous XMM-Newton and Chandra Spectral Analysis \label{simul}} 

Once we had a satisfactory fit with XMM-Newton, we proceeded to model the  contemporary Chandra spectra. We fit the Chandra data applying  a model consisting of
Model C  (including the three ionized absorbers detected in the soft energy band) plus the nine emission lines included in the XMM-Newton fits. Both MEG 
[2 - 25 \AA : 0.5 - 6.2 keV] and HEG data [1.6 - 15 \AA : 0.8 - 7.7 keV] were fitted simultaneously. 

All nine observations (including those from XMM-Newton and Chandra) were fitted simultaneously leaving free to vary independently in each observation the Photon Index 
and normalization of the Power Law, and the normalization of the Black Body. The temperature  kT [keV]  of the Black Body emission was free to vary, but constrained to 
have a single value in all data. The ionization parameter  and the column density of the three absorbing components was also fitted independently among different 
observations. However, as before, the outflow and turbulent velocities of each phase were free to vary, but linked between the nine observations to give a single best-fit 
value.

\begin{figure}
 \includegraphics[width=.57\textwidth]{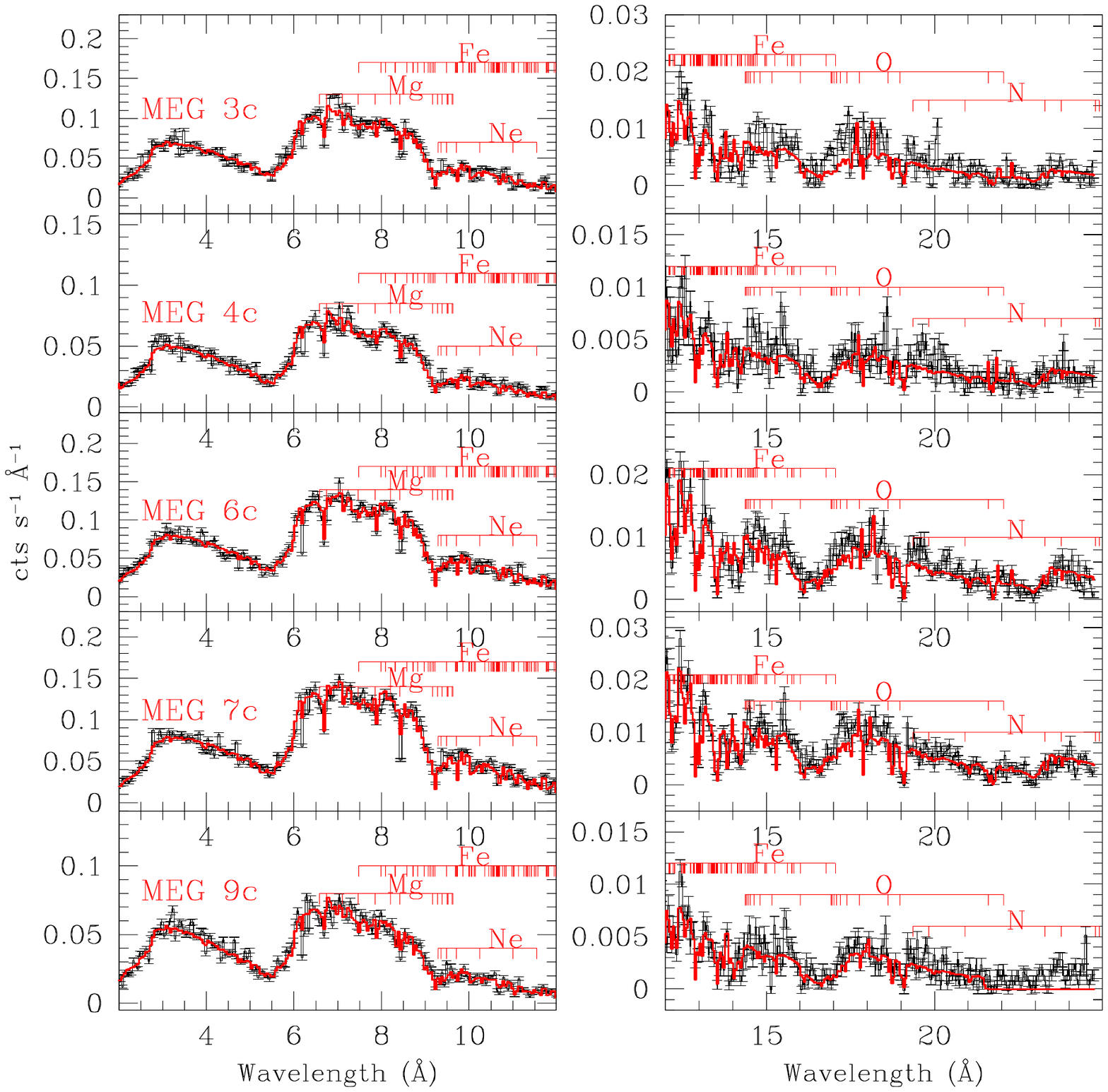}
\caption{MEG spectra in [2 - 25 \AA{}] range of each Chandra observation together with the 
{\bf Model C} fit -red-. The absorption lines are identified above with the corresponding transition 
-red-.\label{fig7}}
\end{figure}

The best fit parameters of this final model  are presented in Table \ref{Table9}. The fits over the high resolution (RGS) spectra are shown in Figures \ref{fig2},\ref{fig3}, 
\ref{fig4} and \ref{fig5} for observations {\bf 1x}, {\bf 2x}, {\bf 5x} and {\bf 8x} respectively.  The MEG of Chandra data and best fit models  are shown in Figure \ref{fig7}. All statistical 
results are satisfactory with $\chi_{red}\sim$1 (Table \ref{Table9}). We note that the values for the ionization parameter and column density on each XMM-Newton observation are 
consistent within 35\% (and within the errors) with those found over the fits excluding the Chandra data (Model C, \S 3.2).

We further included in our models the fourth absorbing component (present only in the hard energy band), leaving free to vary independently among the observations 
the ionization parameter and column density, but constraining the outflow and turbulent velocities to a single best-fit value in all observations.  Our results are summarized in 
Tables \ref{Table9} and \ref{VeryH}. 

Finally, given that the outflow velocities of components {\bf HI} and  {\bf LI} are similar, we tried a model with these parameters linked to a single value for both components. 
The warm absorber best fit parameters did not change significantly than with the Model C, \S 3.2. The velocity found for those two phases (phase {\bf HI} and {\bf LI}) is 
$v_{out} = -1905\pm350$ kms$^{-1}$ .

\begin{deluxetable}{rrrrrrrrr} 
\rotate
\tabletypesize{\scriptsize}
\tablecolumns{7} 
\tablewidth{0pc} 
\tablecaption{{\bf Model C} all XMM-Newton and Chandra Observations\label{Table9}} 
\tablehead{ 
\colhead{\bf Observation} & \colhead{\bf Power Law} &  \colhead{\bf Black Body} & \colhead{Phase {\bf HI}} 
& \colhead{Phase {\bf MI}} & \colhead{Phase {\bf LI}} & \colhead{\bf Statistics} \\
\colhead{} & \colhead{$\Gamma$} & \colhead{kT [keV]} & \colhead{logU} & \colhead{logU} 
& \colhead{logU} & \colhead{$\chi_{red}/dof$}  \\
\colhead{} & \colhead{Norm$^{a}$} & \colhead{Norm$^{c}$} & \colhead{logN$_{H}$} 
& \colhead{logN$_{H}$} & \colhead{logN$_{H}$} & \colhead{} & \colhead{} \\ 
\colhead{} & \colhead{} & \colhead{} & \colhead{v$_{out}$ [kms$^{-1}$]} 
& \colhead{v$_{out}$ [kms$^{-1}$]} & \colhead{v$_{out}$ [kms$^{-1}$]} & \colhead{} \\
\colhead{} & \colhead{} & \colhead{} & \colhead{$-1847\pm350$} & \colhead{$-605\pm350$} 
& \colhead{$-2426\pm350$}  & \colhead{}        } 
\startdata
{\bf 1x} & $1.82\pm0.02$ & $0.09\pm0.02$ & $1.76\pm0.02$ & $0.42\pm0.02$ & $-1.07\pm0.04$ & 
$0.87/4504$ \\ 
 & $1.69\pm0.02$ & $6.47^{+0.06}_{-0.09}$ & $22.33\pm0.02$ & $21.56\pm0.02$ & $21.13\pm0.05$ &  \\ 
 &  &  & $-1847\pm13$ & $-605\pm27$ & $-2426\pm5$ & \\ 
\hline
{\bf 2x} & $1.77\pm0.02$ &  & $1.77\pm0.02$ & $0.36\pm0.02$ & $-0.73\pm0.02$ & $1.01/4558$ \\   
 & $1.33\pm0.02$ & $6.67\pm0.05$ & $22.27^{+0.04}_{-0.02}$ &$ 21.46\pm0.02$ & $21.17\pm0.03$ & \\ 
\hline
{\bf 3c} & $1.509^{+0.033}_{-0.028}$ &  & $2.14^{+0.08}_{-0.09}$ & $0.28^{+0.17}_{-0.12}$ & 
$-0.815^{+0.204}_{-0.207}$ & $0.62/980$ \\   
 & $1.05\pm0.02$ & $4.14\pm0.06$ &$ 22.37^{+0.09}_{-0.08}$ & $21.408^{+0.119}_{-0.082}$ & 
 $21.53^{+0.07}_{-0.08}$ & \\ 
\hline
{\bf 4c} & $1.39^{+0.03}_{-0.02}$ &  & $2.05\pm0.08$ & $0.16^{+0.17}_{-0.19}$ & 
$-0.59^{+0.16}_{-0.17}$ & $0.68/980$ \\   
 & $0.66\pm0.02$ & $2.84\pm0.04$ & $22.22\pm0.09$ & $21.40^{+0.09}_{-0.07}$ & 
 $21.55\pm0.06$ & \\ 
\hline
{\bf 5x} & $1.62\pm0.02$ &  & $1.82^{+0.05}_{-0.08}$ & $0.27\pm0.03$ & $-1.13\pm0.02$ & $0.87/4504$ \\    
 & $0.905^{+0.009}_{-0.011}$ & $4.25\pm0.06$ & $22.49\pm0.12$ & $21.59\pm0.02$ 
 & $21.29^{+0.04}_{-0.02}$ &  \\ 
\hline
{\bf 6c} & $1.58\pm0.02$ &  & $2.24\pm0.03$ & $0.65^{+0.045}_{-0.24}$ & $-0.54\pm0.17$ & $1.08/980$ \\   
 & $1.38\pm0.03$ & $5.65\pm0.04$ & $22.44^{+0.04}_{-0.05}$ & $21.67^{+0.03}_{-0.04}$ & 
 $21.45^{+0.03}_{-0.06}$ & \\ 
\hline
{\bf 7c} & $1.69\pm0.02$ &  & $2.06^{+0.07}_{-0.03}$ & $0.53^{+0.04}_{-0.08}$ & 
$-0.65^{+0.13}_{-0.17}$ & $0.95/980$ \\    
 & $1.58^{+0.05}_{-0.04}$ & $6.13\pm0.05$ & $22.40\pm0.05$ & $21.61^{+0.04}_{-0.06}$ & 
 $21.403^{+0.039}_{-0.077}$ & \\ 
\hline
{\bf 8x} & $1.75\pm0.02$ &  & $2.03\pm0.02$ & $0.45\pm0.02$ & $-0.81\pm0.04$ & 
$0.95/4630$ \\   
 & $1.41\pm0.02$ & $5.33\pm0.02$ & $22.52\pm0.02$ & $21.47\pm0.02$ & $21.31\pm0.02$ &  \\ 
\hline
{\bf 9c} & $1.27^{+0.03}_{-0.02}$ &  & $2.45^{+0.05}_{-0.12}$ & $0.17\pm0.15$ & 
$-0.83^{+0.22}_{-0.26}$ & $0.74/980$ \\    
 & $0.62^{+0.03}_{-0.02}$ & $2.94\pm0.05$ & $22.44^{+0.08}_{-0.09}$ & $21.57\pm0.06$ & 
 $21.54^{+0.06}_{-0.07}$ & \\
\enddata 
\tablenotetext{a}{In $10^{-2}$ $photons/keV/{cm^{2}}/s$ in at 1 keV.}
\tablenotetext{c}{In $10^{-4}$ $L_{39}/D_{10}^{2}$.} 
\tablenotetext{d}{In $10^{-11}$ $ergs/cm^{-2}/s$.} 
\end{deluxetable} 

\subsection{Fitting the two absorption features in the hard X-ray band of XMM-Newton and Chandra} 

\begin{figure}
 \includegraphics[width=0.6\textwidth]{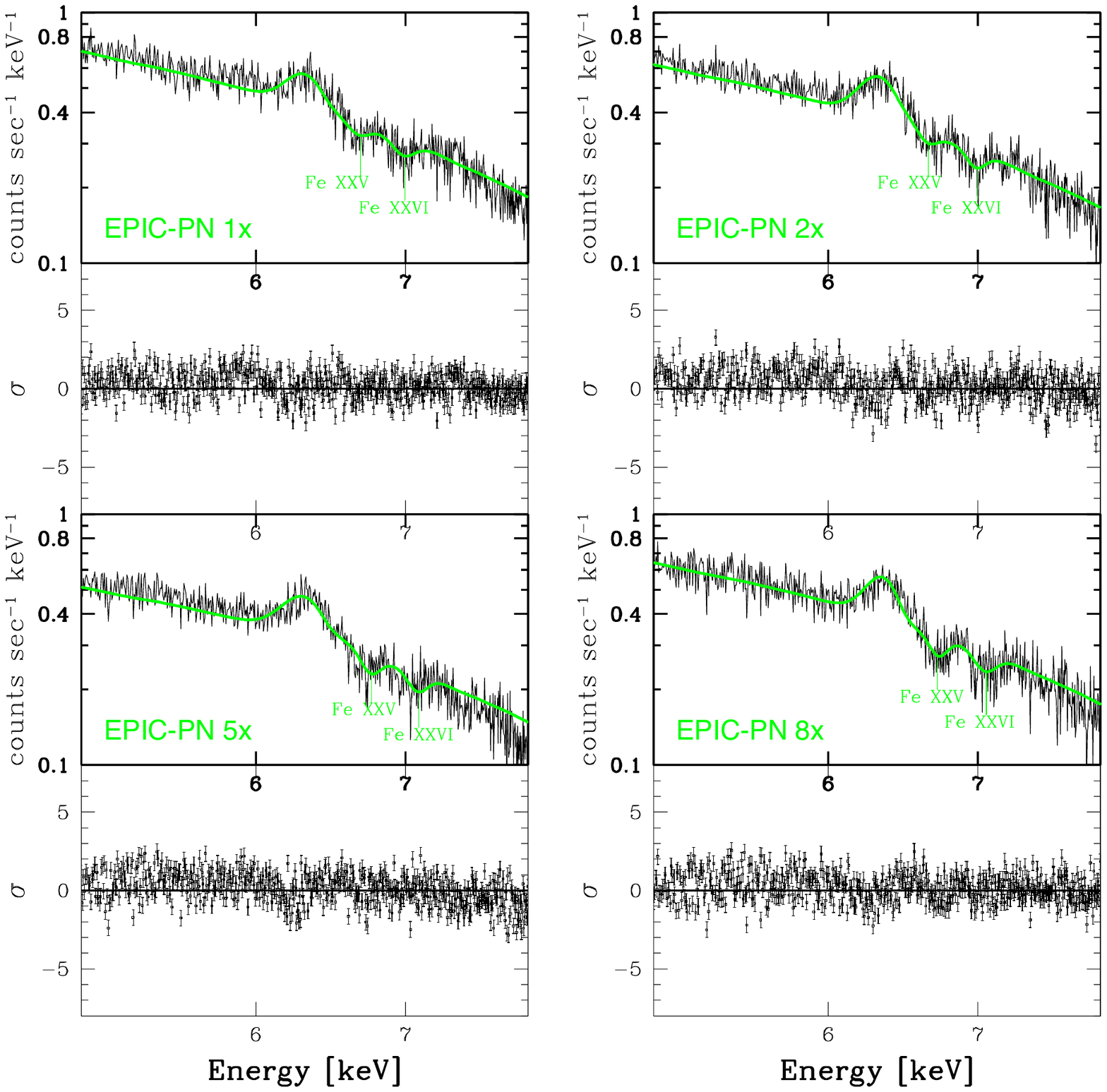}
\caption{EPIC-PN spectra in [5 - 8 keV] range of each XMM-Newton observation together 
with the Power Law function absorbed by phase {\bf VH} fit (in green).\label{VH_abs}}
\end{figure}

We included in our models a fourth WA component with a very high ionization level, to fit the two absorption lines detected in the hard band (these lines 
were initially identified with transitions by FeXXV and FeXXVI and fitted with Gaussians over the EPIC-PN data). This fourth absorber was modelled only on the EPIC-PN 
data of XMM-Newton from 3 to 10 keV and the HEG (High Energy Grating) spectra of Chandra in the 3 - 7.7 keV hard X-ray band. An F-test gives us a confidence level l
arger than $99.99\%$ for the existence of this absorber (hereafter {\bf VH}) in all XMM-Newton observations. We found logU $\sim$ 3.8 and logN$_{H}\sim$ 23.2 (see 
Table \ref{VeryH}). Figure \ref{VH_abs} shows the spectra and best fit over the EPIC-PN data.

It is important to clarify that the hard X-rays were fitted separately of the soft X-ray band. However, to fix the power law in the soft energy band we took account into the 
hard band in the fit, with the aim to have the best fit value connected with the soft energy band. We also note that the soft absorbers have no effect on the hard energy band.

First, we fitted simultaneously three spectra of each XMM-Newton observation, RGS 1 \& RGS 2 [8 - 38 \AA:0.33 - 1.55 keV] and EPIC-PN spectra in the energy range of 0.3 - 10 
keV, the EPIC-PN spectra were included with the aim to obtain the best parameter of the power law. Then, we fitted the Fe-K$\alpha$ emission line in 6.38 keV and two absorption 
lines of FeXXV and XXVI in 6.7 and 6.96 keV respectively. In this model (MODEL C) were included 10 emission lines in the soft energy band. The model C was applied 
successfully into the five spectra of  MEG [2 - 25 \AA:0.5 - 6.2 keV] and HEG spectra [1.6 - 15 \AA:0.8 - 7.7 keV], both Chandra high resolution detectors. Finally we cut the spectra of 
EPIC-PN from 3 to 10 keV and HEG spectra from 3 - 7.7 keV to obtain the fourth and highest ionized absorber VH.

\begin{deluxetable}{cccc} 
\tabletypesize{\scriptsize}
\tablecolumns{4} 
\tablewidth{0pc} 
\tablecaption{Phase {\bf VH} applied in the hard band spectra [3 - 10 keV], EPIC-PN of 
XMM-Newton and HEG [3 - 7.7 keV] of Chandra. The outflow velocity is given by v$_{out}$=-2650$\pm$350 
kms$^{-1}$.\label{VeryH}} 
\tablehead{
\colhead{\bf Obs} & \colhead{\bf VH.logU} & \colhead{\bf VH.logN$_{H}$} 
& \colhead{\bf Statistics}  \\ 
\colhead{} &  \colhead{} & \colhead{}  & \colhead{$\chi_{red}$/dof} }
\startdata
{\bf 1x}  & 3.87$^{+0.09}_{-0.08}$ & 23.22$\pm$0.02   & 0.65/1067  \\ 
{\bf 2x} & 3.88$^{+0.08}_{-0.06}$ & 23.22$\pm$0.02   & 0.81/1116  \\ 
{\bf 3c} & 3.99$\pm...$ & 23.26$^{+0.07}_{-0.24}$ &   0.76/103  \\ 
{\bf 4c} & 3.91$^{+...}_{-0.32}$ & 23.22$\pm$0.06 &    1.25/103  \\ 
{\bf 5x} & 3.88$\pm$0.42 & 23.23$\pm$0.02 &   0.78/1062  \\ 
{\bf 6c} & 4.14$^{+...}_{-0.06}$ & 23.27$^{+0.03}_{-0.59}$ &  1.33/103  \\ 
{\bf 7c} & 3.51$^{+...}_{-0.91}$ & 23.18$^{+0.02}_{-0.52}$ & 1.01/103  \\ 
{\bf 8x} & 3.75$\pm$1.08 & 23.22$\pm$0.02 & 0.7/1188  \\ 
{\bf 9c} & 3.49$^{+1.27}_{-0.37}$ & 23.28$^{+0.27}_{-0.37}$ & 0.82/103  \\ 
\enddata 
\tablenotetext{*}{{\bf ...} means indeterminate.}
\end{deluxetable} 

\section{Summary of Results and Comparison with Previous Works\label{results}}
 
In order to study in detail the variability of the WA in NGC 3516, we have re-analyzed nine different individual observations  from XMM-Newton and Chandra satellites. Our 
results show that the best fit model for all spectra consists of a variable continuum emission absorbed by three partially ionized absorbers (WA) accounting for the soft band 
and a fourth highly ionized WA detected only in the hard band. Tables \ref{Table9} and \ref{VeryH} summarize the values of the parameters of this model (Model C) for 
each observation. 
As can be observed in these tables, there are significant variations in some parameters of the model among observations. In the following, we discuss the general properties of 
these absorbers, and defer the time resolved analysis for section \ref{variability}.

\begin{figure}
 \includegraphics[width=0.6\textwidth]{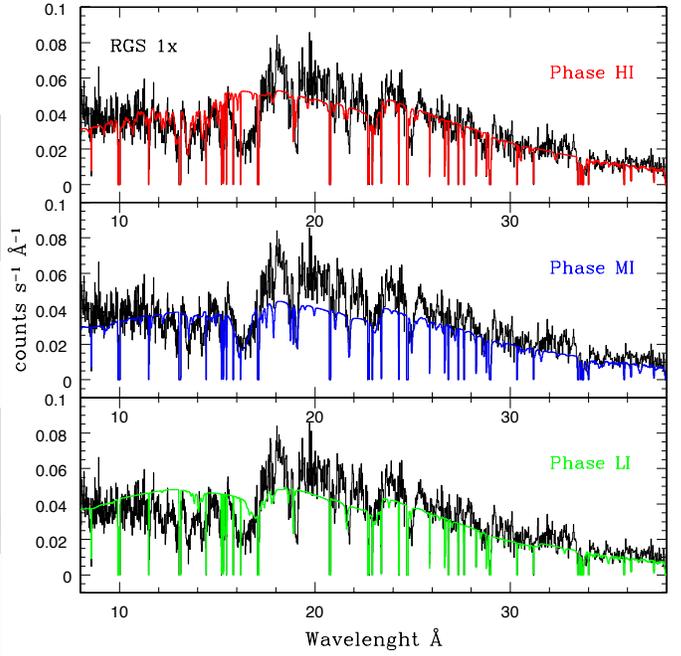}
\caption{RGS spectrum of observation {\bf 1x} -black- from 8 to 38 \AA{}. Phase {\bf HI}  
(top panel) is drawn in red color, as it can be observed, it reproduces absorption lines mainly in the 
lowest wavelengths of the spectrum. Phase {\bf MI}  (medium panel), in blue color, reproduces the Fe 
M-shell UTA array and some absorption lines due to OVII and OVIII transitions. Finally, phase {\bf LI}  
(bottom panel), drawn in green color, generates a fraction of the Fe-UTA and also the main absorption 
features around 18 \AA{}.\label{fig8}}
\end{figure}

The different ionization components produce different absorption features. If we analyze them we found that, in the soft band, the highest ionized phase {\bf HI} produces 
absorption by several Fe L-shell transitions with charge state XVIII - XXII in the 10 - 16 \AA{} range; absorption features by SiXIII - SiXIV (between 5 and 7 \AA); NeX 
absorption lines at 9.29, 9.36, 9.48 and 10.24 \AA; and the OVIII transition at 18.97 \AA. Phase {\bf HI} electron temperature is  T$_{HI,e}=4.5\times10^{5}$ K and the outflow 
velocity is v$_{out,HI}=-1847\pm350$ kms$^{-1}$. Phase {\bf MI} imprints absorption features due to OVII (at 16.96, 17.01, 17.09, 17.2, 17.4, 17.77, 18.62 and 21.6 \AA{}), OVIII 
(at 14.63, 14.45, 14.41, 14.82, 18.97 \AA{}), and the most energetic portion of the Fe M-shell UTA (between 15 and 17 \AA{}). The electron temperature associated to phase 
{\bf MI} is given by T$_{MI,e}=9.8\times10^{4}$ K, while the outflow velocity is v$_{out,MI}=-605\pm350$ kms$^{-1}$. Phase {\bf LI} produces absorption lines by OVII at $\sim$ 16.9 - 
18.6 \AA{} and the low energy part of the M-shell Fe UTA (16.5 - 18 \AA{}). This phase also produces absorption lines by NeV - NeVI between 12 and 14 \AA{}. It has an
electron temperature  T$_{LI,e}= 3.0 \times 10^{4}$ K and an outflow velocity v$_{out,LI}=-2426\pm350$ kms$^{-1}$. Figure \ref{fig8} shows the absorption features produced 
by each of the three absorbing phases ({\bf HI}, {\bf MI} and {\bf LI}). 

On other hand, the fourth hard absorption phase {\bf VH} produces the Fe K complex in the hard band. Its electronic temperature and outflow velocity are
 T$_{VH}= 7.0\times10^{5}$ K and v$_{out,VH}=-2649$ kms$^{-1}$ respectevely (see Table \ref{Table9}). 

Given that several works have analyzed the 2006 X-ray spectra of this source, a comparison among different results is mandatory. In Table \ref{comparison} we present the 
different parameters found in the different analyses. We note that the definition of the ionization parameter used by \cite{tur08} and \cite{meh10}, is different than the one 
presented here. These authors used the ionization parameter $\xi$ \footnote{Ionization parameter $\xi$ : $\xi \equiv L/nr^{2}$ where $L$ is 1-1000 Rydbergh luminosity, 
$n$ the gas density and $r$ the absorber-source distance.}, while our analysis was done  with the parameter U $^{1}$. Also, the Spectral Energy Distribution is different in the 
different analyses. Therefore, a direct comparison of the ionization parameter is not possible. However, all works are in agreement  in the sense that all require three different 
ionized absorbers, each imprinting absorption by similar charge states in the soft X-ray energy band. The fourth highly ionized absorber (found in the hard energy band) 
reported by \cite{tur08} as {\bf Zone 4} clearly corresponds with phase {\bf VH} in this work.

In general, the column densities are also in agreement among the different works. The column density N$_{H}$ of phase {\bf LI} is similar to that reported by \cite{tur08} and  
\cite{meh10}. The values of N$_{H}$ for components {\bf HI} and {\bf VH} correspond to those by \cite{tur08} for {\bf Zone 3} and {\bf Zone 4}, respectively. We note, however, 
that there is a significant difference in the column density of the mid-ionized absorber {\bf MI} with  respect to the {\bf Zone 2} \citep{tur08} and {\bf Phase B} 
\citep{meh10}. This change can be explained by the fact that we assumed full coverage for all absorbers in our analysis (but see \S \ref{coverage}). 

The most intriguing difference in this work with respect to previous analyses is present in the different outflow velocities of the absorbers: Phases {\bf LI}, {\bf HI} and {\bf VH} 
present outflow velocities significantly larger than previously found. Instead, phase {\bf MI} presents a smaller outflow velocity than the components of middle ionization of these 
other works. It is worth to mention that in \cite{tur05} the velocity of their {\bf Zone 1} was fixed to an outflow velocity of -200 kms$^{-1}$, but in \cite{tur08} a value v$_{out}\sim$ 
-1000 kms$^{-1}$ is reported for this same component. Furthermore, \cite{tur08} based the outflow velocities in only 11 absorption lines (see their Table 1). Overall, the 
differences in velocity are likely produced because the different absorption lines are present with a range of velocities, that are not fully resolved in the data. 
 
This is further evidenced by \cite{hol12}, who made a detailed analysis of the kinematic components in the high resolution spectra of Chandra: They found evidence of four 
kinematic components with different outflow velocities. According to these authors, the slower components (with v$_{out}$=-350$\pm$100 kms$^{-1}$ and 
v$_{out}$=-1500$\pm$150 kms$^{-1}$) present a broad range of ionization, while the two faster components (v$_{out}$=-2600$\pm$200 kms$^{-1}$ and 
v$_{out}$=-4000$\pm$400 kms$^{-1}$) are more highly ionized. Phases {\bf MI}, {\bf HI} and {\bf VH} of the model proposed in this work are consistent with the results by 
\cite{hol12}. The higher outflow velocity of component {\bf LI} is discussed below, in section \ref{multiph}.

\begin{deluxetable}{lcccc} 
\rotate
\tabletypesize{\scriptsize}
\tablecolumns{5} 
\tablewidth{0pc} 
\tablecaption{Comparison between the warm absorbers found in \citealp{tur08}, \citealp{meh10} and this work \label{comparison}.} 
\tablehead{ 
\colhead{\bf Model} & \colhead{\bf WA 1} & \colhead{\bf WA 2} & \colhead{\bf WA 3}  & \colhead{\bf WA 4} \\}
\startdata
{\bf Turner} & {\bf Zone 1} & {\bf Zone 2} & {\bf Zone 3} & {\bf Zone 4} \\   
& log$\xi\sim$ -0.05 & log$\xi \sim$ 3 &  log$\xi\sim$ 2.5 & log$\xi\sim$ 4.3  \\ 
& N$_{H}\sim$ 6$\times$10$^{21}$ cm$^{-2}$ & N$_{H}\sim$10$^{22}$ cm$^{-2}$ & N$_{H}\sim$10$^{23}$ cm$^{-2}$ & N$_{H}\sim$10$^{23}$ cm$^{-2}$ \\ 
& v$_{out}$ =-200 kms$^{-1}$ & v$_{out}$ =-1100 kms$^{-1}$ & v$_{out}$=-1100 kms$^{-1}$ & v$_{out}$=-1000 kms$^{-1}$ \\ 
\hline
{\bf Mehdipour} & {\bf Phase A} & {\bf Phase B} & {\bf Phase C} &  \\ 
& log$\xi$=[0.87-0.97$\pm$0.02] & log$\xi$=[2.39-2.43$\pm$0.05] & log$\xi$=[2.99-3.07$\pm$0.05] &  \\ 
& N$_{H}$=[0.33-0.43 $\pm$ 0.02]$\times$10$^{22}$ cm$^{-2}$ & N$_{H}$=[1.7-3.2 $\pm$ 0.04]$\times$10$^{22}$ cm$^{-2}$ & 
N$_{H}$ = [1.2-1.7$\pm$ 0.02]$\times$10$^{22}$ cm$^{-2}$ &  \\ 
& v$_{out}$ = -100 to -200 $\pm$ 40 kms$^{-1}$ & v$_{out}$ = -1500 to -1600 $\pm$ 100 kms$^{-1}$ & v$_{out}$ = -800 to -1000 $\pm$ 300 kms$^{-1}$ &  \\ 
\hline
{\bf Huerta} & {\bf Phase LI} & {\bf Phase MI} & {\bf Phase HI} & {\bf Phase VH} \\ 
& logU=-1.07 to -0.54$\pm$0.17 & logU=[0.16-0.65]$pm$0.04 & logU=[1.76-2.45]$pm$0.05 & logU=[3.49-4.14]$pm$0.06 \\ 
& N$_{H}$=[0.13-0.34 $\pm$0.01]$\times$10$^{22}$ cm$^{-2}$ & N$_{H}$=[0.25-0.40 $\pm$0.01]$\times$10$^{22}$ cm$^{-2}$ & 
N$_{H}$=[1.66-3.31$\pm$0.01]$\times$10$^{22}$ cm$^{-2}$ & N$_{H}$=[1.51-1.90$\pm$0.01]$\times$10$^{23}$ cm$^{-2}$ \\ 
& v$_{out}$ = -2426$\pm$350 kms$^{-1}$ & -605$\pm$350 kms$^{-1}$ & -1847$\pm$350 kms$^{-1}$ & -2650$\pm$350 kms$^{-1}$ \\ 
\hline
\enddata 
\tablenotetext{*}{Ionization parameter $\xi \equiv L/nr^{2}$ where $L$ is 1-1000 Rydbergh luminosity, $n$ de gas density and $r$ the absorber-source distance.}
\tablenotetext{*}{Ionization parameter $U \equiv \frac {Q}{4{\pi}R^{2}n_{H}c}$.}
\end{deluxetable} 

\subsection{Time Variability \label{variability}}

In this section, we perform a time resolved analysis of the ionized absorber in NGC 3516. In particular, we identify the parameters that show significant variations among the 
observations and explore a possible correlation with the changes in flux. 

\subsubsection{Ionization Parameter U \label{vh}}

\begin{figure}
 \includegraphics[width=0.55\textwidth]{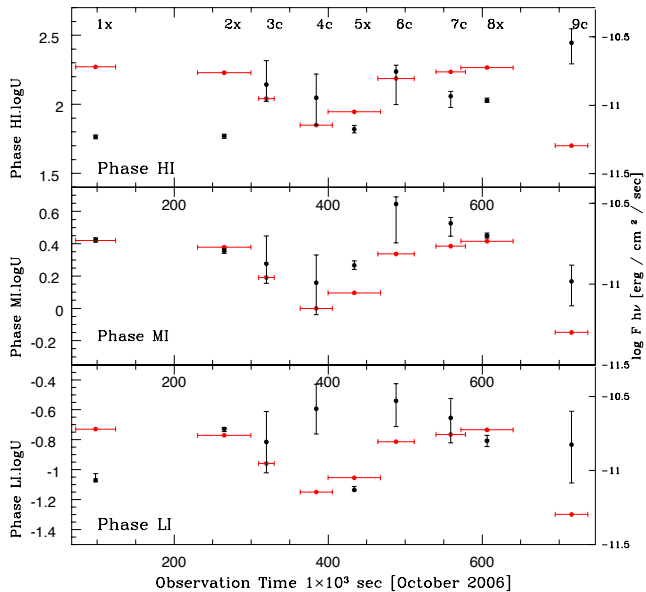}
\caption{In all panels is plotted the light curve of the X-ray [8 - 25 \AA{}] range of NGC 3516 
-red-, the flux energy F$_{h\nu}$ was scaled in order to compare it with the logU parameter of each 
ionized phase. The logU variation in time is plotted for phases {\bf HI} (top panel), {\bf MI} 
(medium plot) and {\bf LI} (bottom plot).\label{fig9}}
\end{figure}

{\bf Phase HI}: The most significative change in ionization parameter is given at a level 4$\sigma$ between observation {\bf 8x} and the others. This parameter, along the nine 
observations, is plotted in Figure~\ref{fig9} (top plot). We also plot a scaled version of the flux values. This scaled version of the light curve represents the expected changes of 
U if the gas were instantly in photoionization equilibrium with the ionizing source (for gas in photoionization equilibrium, changes in flux must be followed by changes by the 
same factor in U). Clearly, this is not observed in the Figure \ref{fig9}, indicating that there is no clear response of this component to the flux variations. 

{\bf Phase MI}: This component presents strong significative changes between observations {\bf 1x} and {\bf 5x} at 7$\sigma$ and between observations {\bf 5x} to {\bf 8x} at 
$8\sigma$. Interestingly, these changes are correlated with the flux variations. Among observations {\bf 1x}, {\bf 2x}, {\bf 3c} and {\bf 4c}, the ionization degree of the phase {\bf 
MI} changes by nearly the same factor as the X-ray flux. This indicates that this component is responding to the continuum changes close to photoionization equilibrium. 

{\bf Phase LI}: The most important variations in U are found between observations {\bf 1x} and {\bf 4c} and between observations {\bf 2x} and {\bf 5x}, both at $4\sigma$. For 
this phase, there is a reaction to the larger changes in flux in longer timescales, as observed in Figure \ref{fig9} (bottom panel). From observations {\bf 2x} and {\bf 4c} the gas 
presents a delay in recombining as expected in photoionization equilibrium ({\it PhE}). Yet, at the time of observation {\bf 5x}, the gas seems to have found equilibrium again. 
As the flux increases from observations {\bf 5x} to {\bf 6c} the gas re-ionizes to reach an ionization state consistent again with PhE. Overall, this component also seems to 
respond, albeit with larger delays and in a more complex way to the changes in the continuum.

Finally, we note that the highest ionization component {\bf VH} (detected only in the hard X-ray band) does not presents significant changes in the ionization parameter. This 
behaviour is expected because the ionization potential of FeXXV is 8.8 keV and that of FeXXVI is 9.3 keV, and the hard X-rays always vary much less compared to soft X-ray 
band. Therefore, it is possible that at these relevant energies, the continuum did not vary much. We conclude that the lack of variability cannot be used to use a limit on density 
for this component.

\subsubsection{\bf Equivalent H Column Density}

\begin{figure}
  \includegraphics[width=0.55\textwidth]{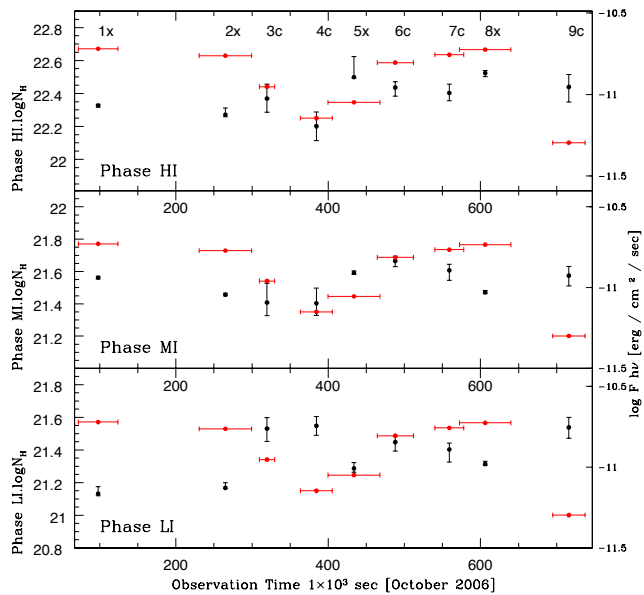}
\caption{The logN$_{H}$ variation in time is plotted for phases {\bf HI} (top panel), {\bf MI} 
(medium plot) and {\bf LI} (bottom plot). The flux energy F$_{h\nu}$ (in red) was scaled in order to compare it with the logN$_{H}$ parameter of each 
ionized phase. \label{fig10}}
\end{figure}

There are mild variations (by a factor $\lesssim$ 2.5) in the H equivalent column density along the nine observations, for the three phases detected in the soft X-ray band. The 
variations are not anti-correlated with the energy flux, as it would be expected in a scenario where the WA consists of material crossing our line of sight, and this material were 
responsible for producing the flux variation observed in the light curve. Figure \ref{fig10} shows the changes in this parameter of each ionized phase (phases {\bf HI}, {\bf MI} 
and {\bf LI}). 

In order to study how significant are the changes in column density, we performed a test, fitting all datasets with a single value of N$_{H}$. The best fit value for this parameter
is similar to the values obtained when modelling individual observations, as expected. However, it is noteworthy that the other parameters are fully consistent with those 
obtained in Model C. In particular, the values for the ionization parameter are the same, within errors, with those found if N$_{H}$ is free to vary in each observation. This 
implies that the variations found in U are not produced or correlated by possible variations in the column density.

\subsection{Variations in Spectral Features in the spectra of NGC 3516}

Given that the models predict variations in the ionization degree of two ionization phases ({\bf MI} and {\bf LI}), we have studied in detail the spectra of the source to identify 
clear signatures of such variations. 

\subsubsection{The Iron M-Shell Unresolved Transition Array}

\begin{figure}
\begin{center}
\begin{tabular}{cc}
\includegraphics[width=1.5in]{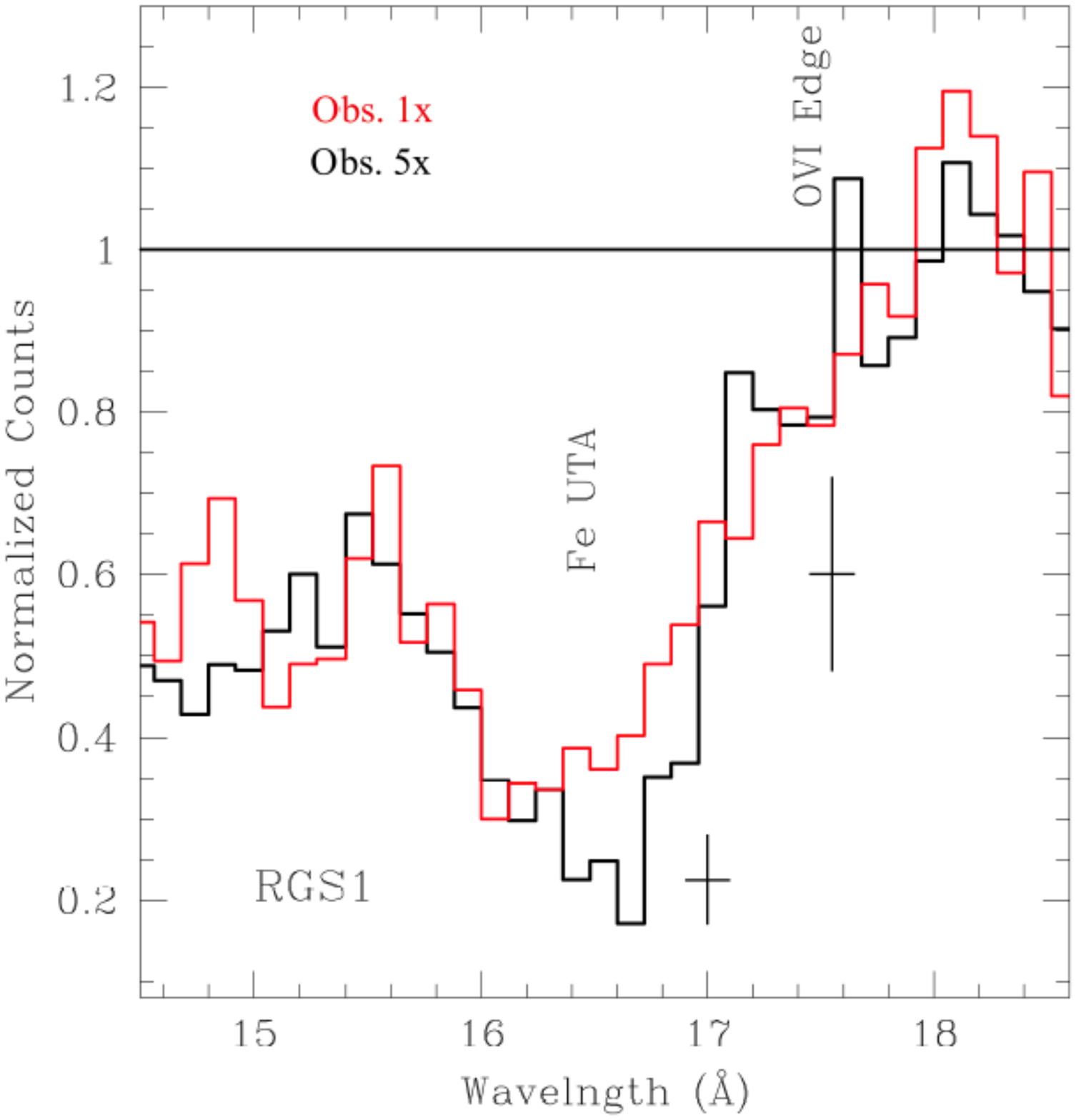}   &
\includegraphics[width=1.5in]{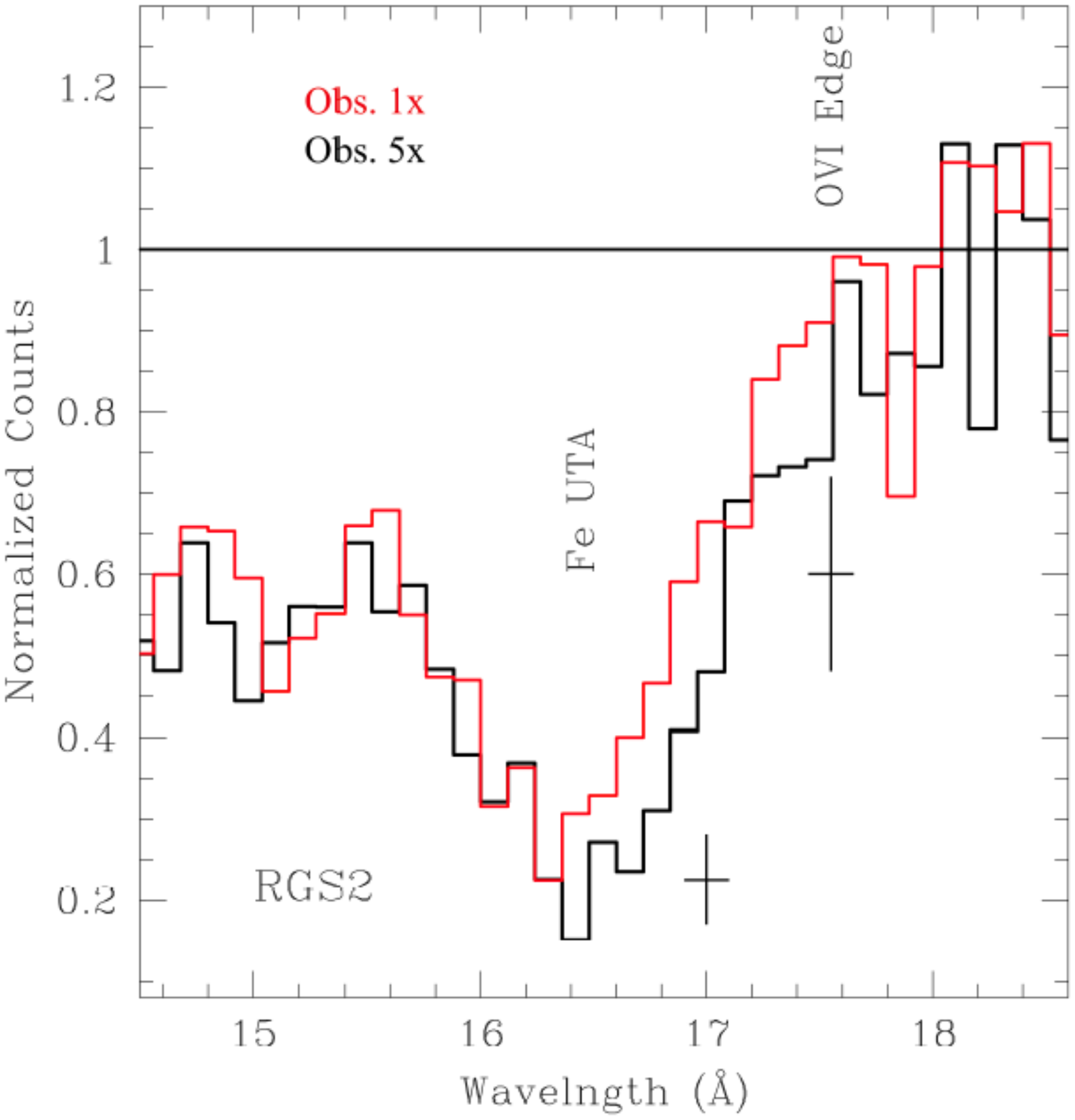}  \\
\includegraphics[width=1.5in]{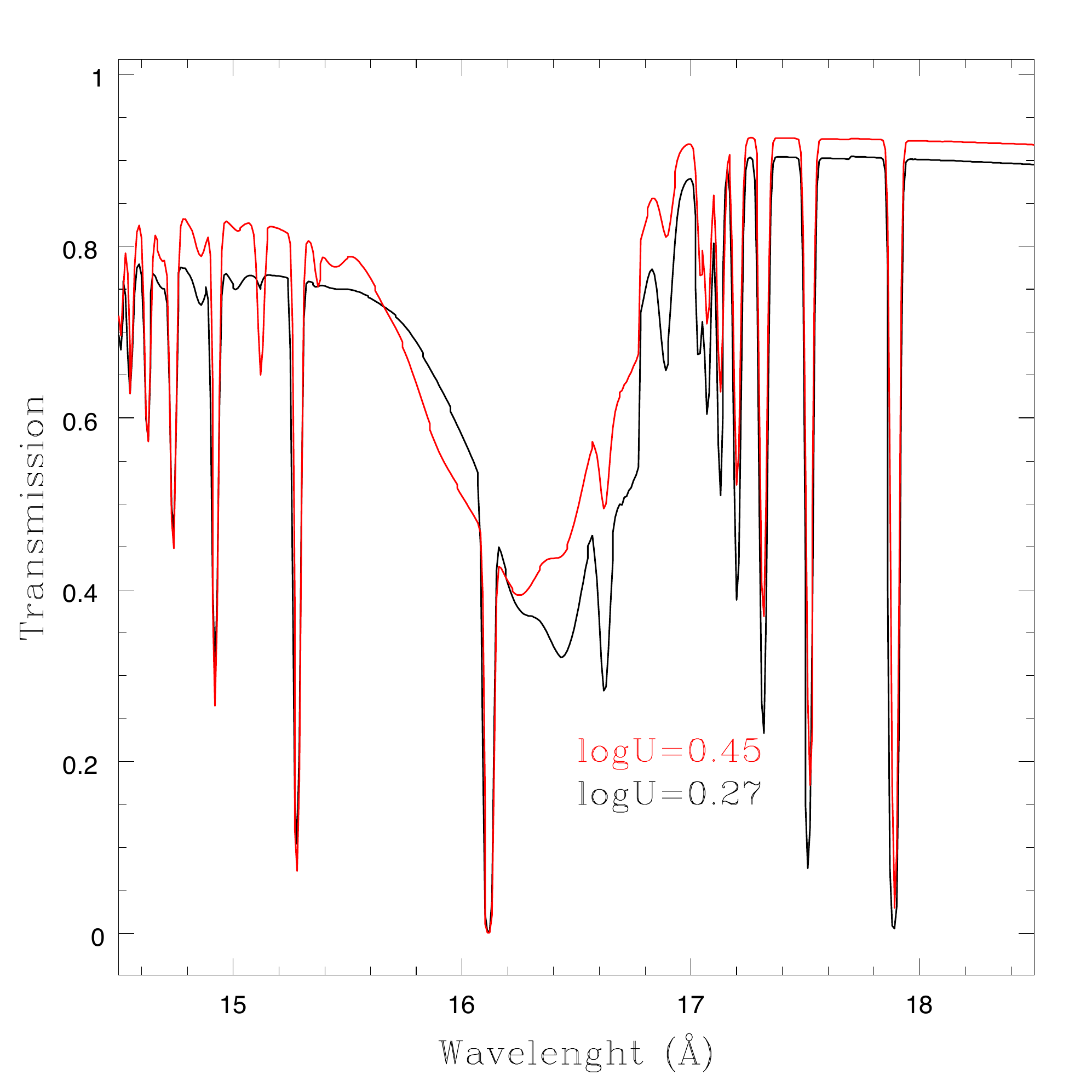} & 
\end{tabular}
\end{center}
\caption{Fe M-shell UTA comparison between {\bf 1x} and {\bf 5x} spectra of RGS from 
XMM-Newton. For RGS 1 top panel left and for RGS 2 top panel right [14.5 - 18.5 \AA{}].
In the bottom panel is plotted the opacity variation produced by a change in the ionizing flux by a factor of 2 in photoionization equilibrium 
(modelled with the {\it PHASE} code \citep{kro03}).
\label{fig11}}
\end{figure}

This Fe M-shell {\it UTA} (Unresolved Transition Array) is extremely useful to look for opacity changes in the absorber phases (as shown by \cite{kro05a}). To look for these 
changes, we performed a comparison between spectra {\bf 1x} and {\bf 5x} in the 15 - 18 \AA{} range. Both spectra were normalized using a simple power law to the local 
continuum. The residuals were over-plotted to search for possible differences. The results are shown in Figure \ref{fig11}. An evident variation in the position of the Fe UTA 
can be observed between spectra {\bf 1x} and {\bf 5x}  (upper panels in Fig. \ref{fig11}). 

In the bottom panel of Figure \ref{fig11} we present the expected opacity changes in this spectral region, in response to a variation by a factor of $\sim$ 2 in the ionizing flux. 
This change is similar to the flux variation between observations {\bf 1x} and {\bf 5x}, so it closely matches an scenario were the gas is in photoionization equilibrium. It is clear 
that the observed variations are consistent with those expected in photoionization equilibrium for components {\bf MI} and {\bf LI}, that produce the UTA.

\begin{figure}
\begin{center}
\begin{tabular}{cc}
\includegraphics[width=1.6in]{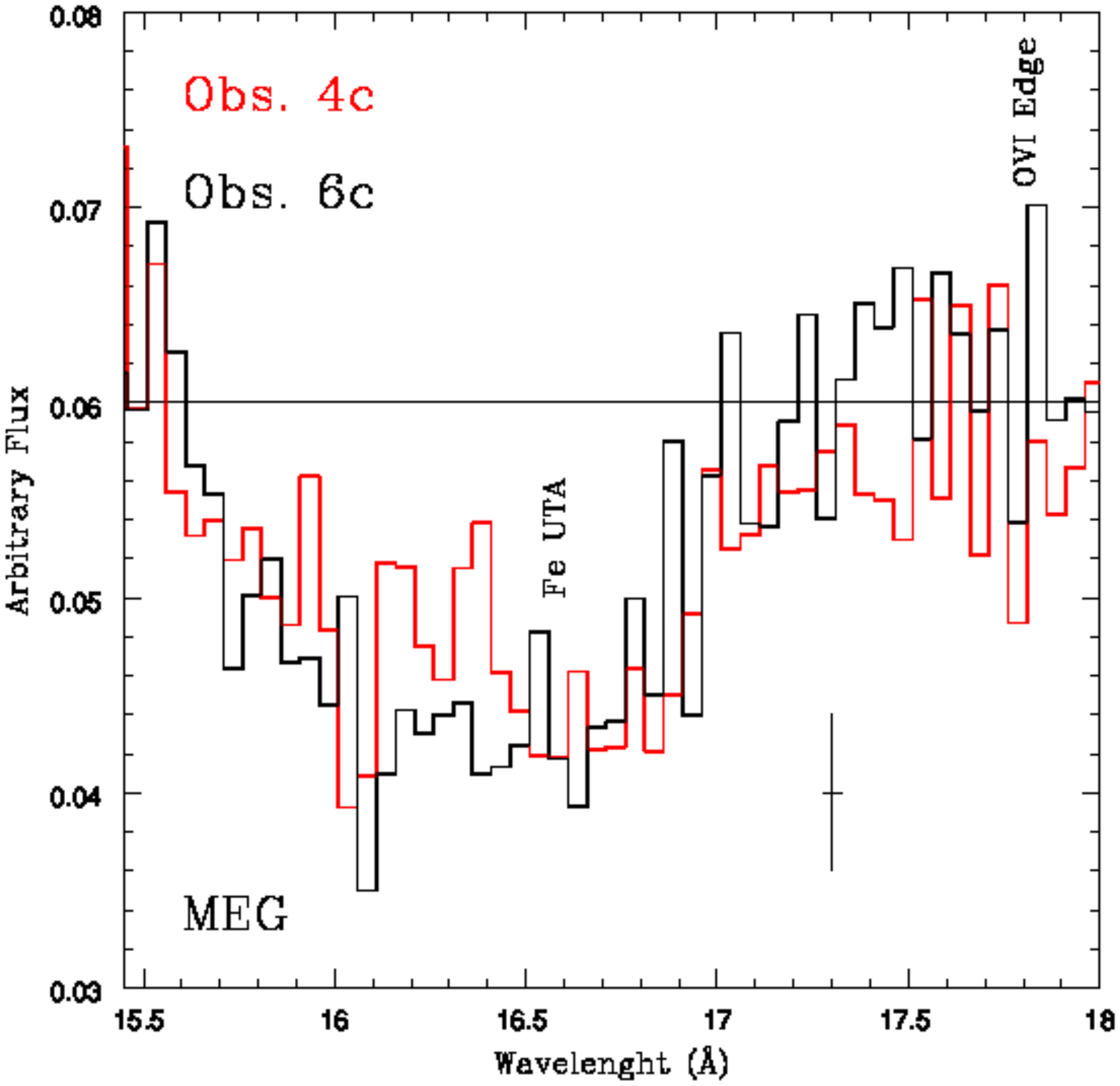}  & 
\includegraphics[width=1.6in]{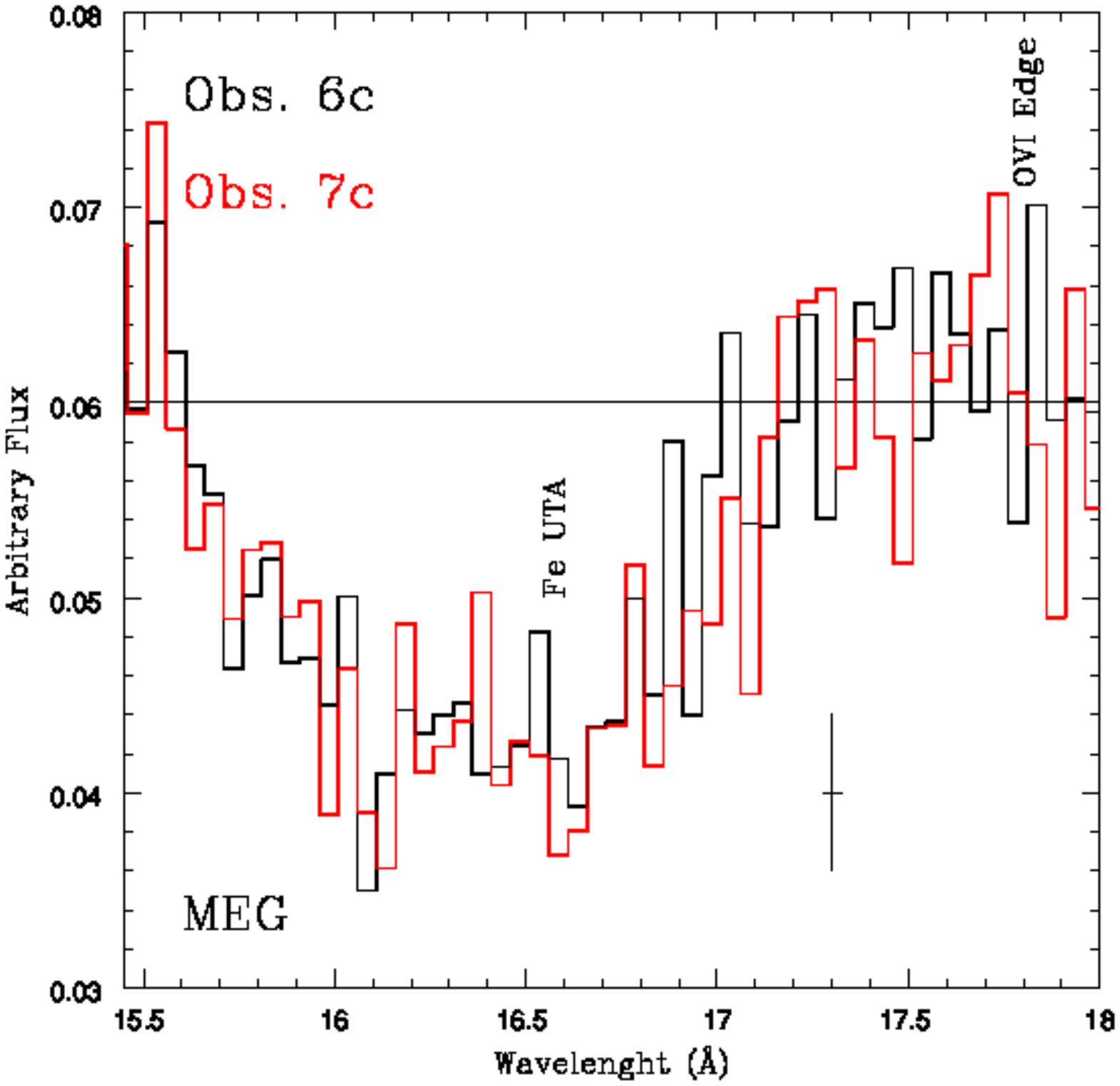} 
\end{tabular}
\end{center}
\caption{
Fe M-shell UTA comparison between Chandra data from 15.5 to 18.5 \AA{} range. Left panel: observations {\bf 4c} and {\bf 6c}. Right panel: observations {\bf 6c} and {\bf 7c}. 
\label{fig12}}
\end{figure}

Figure \ref{fig12} compares the Fe M-shell UTA between different Chandra spectra. This plot shows a similar variability trend (although less significant) between observations 
{\bf 4c} and {\bf 6c}, again pointing to gas close to photoionization equilibrium. Furthermore, no spectral variations are found between observations {\bf 6c} and {\bf 7c}, where 
the flux remains constant.
 
\subsubsection{Testing Possible Changes in Covering Factor \label{coverage}}

\begin{figure}
\begin{center}
\begin{tabular}{cc}
\includegraphics[width=1.6in]{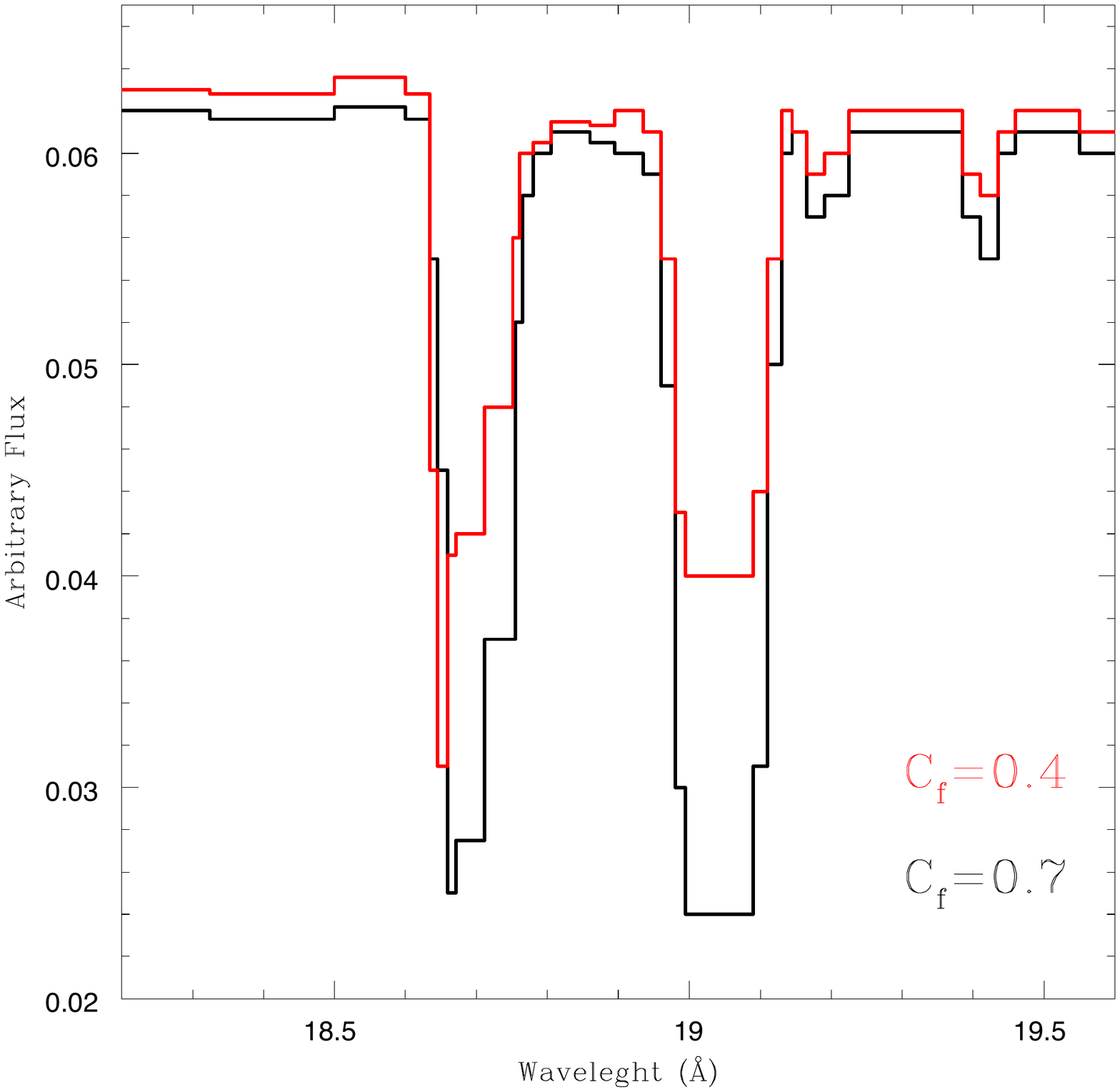} \\
\includegraphics[width=1.6in]{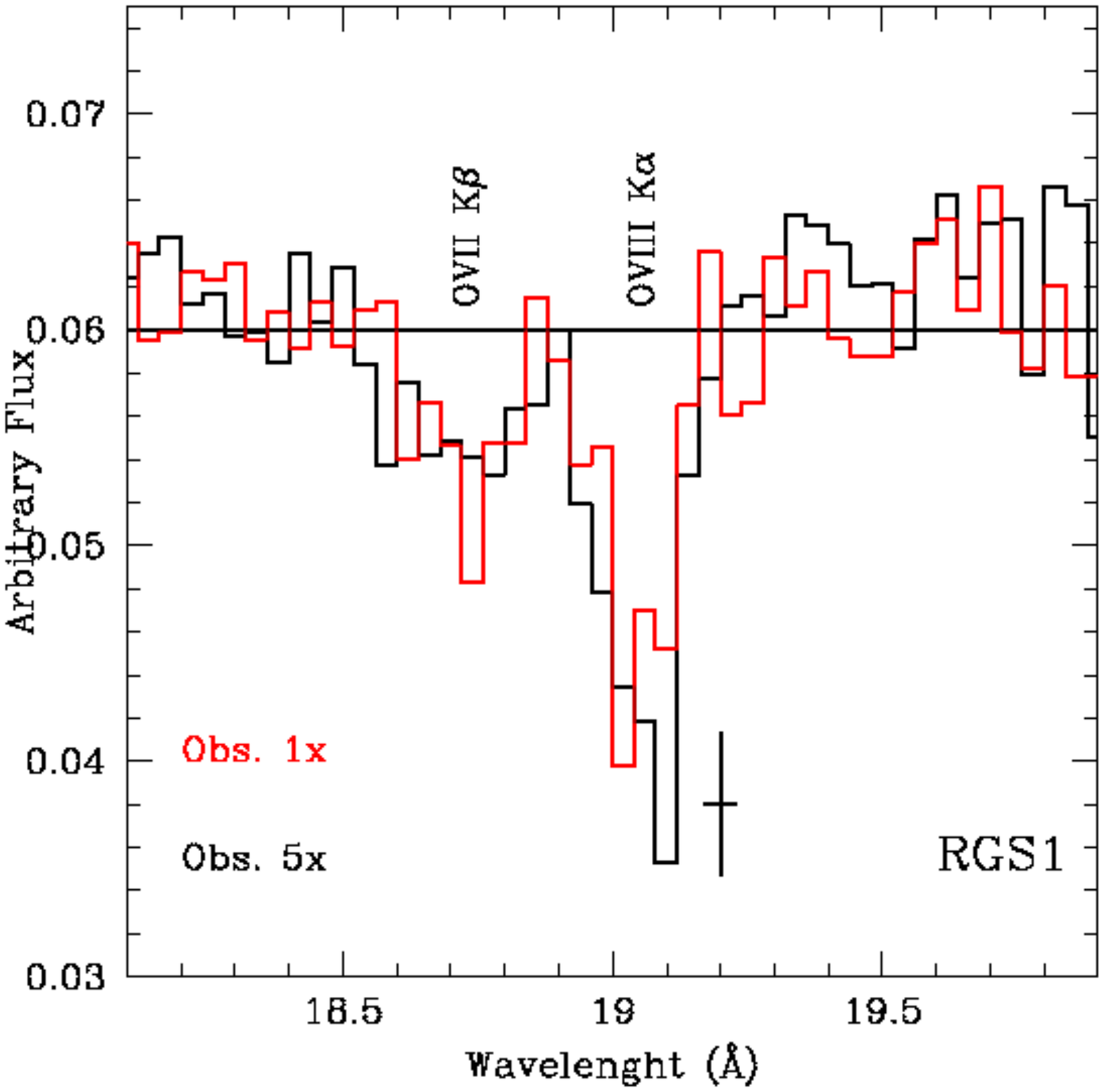}  &  
\includegraphics[width=1.6in]{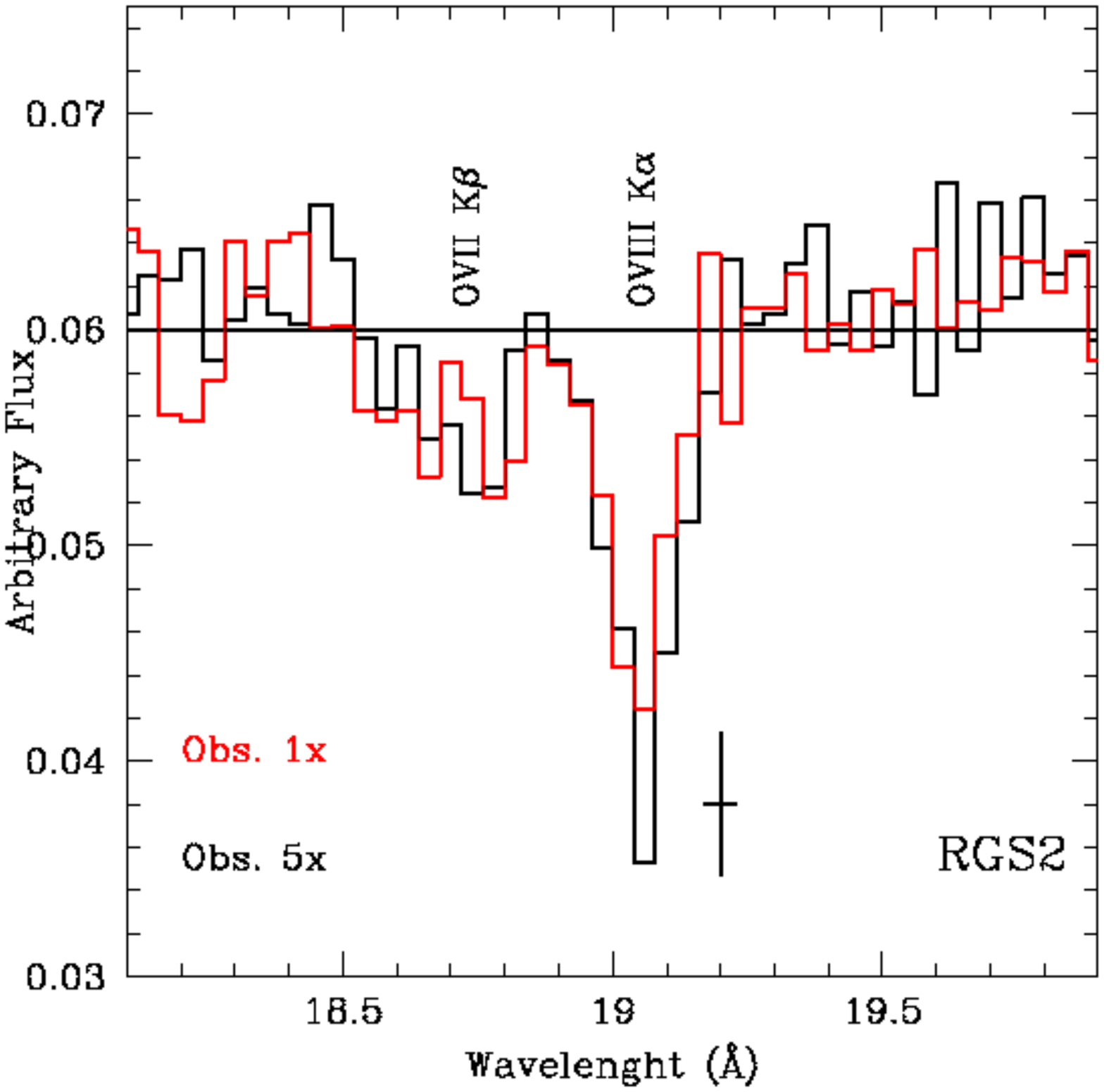} \\
\end{tabular}
\end{center}
\caption{OVII-K$\beta$ and OVIII-K$\alpha$ absorption lines [18 - 20 \AA], the first panel (top) is the model assuming that the covering factor is changing. 
The second (bottom left) and third (bottom right) panels contain the OVII-K$\beta$ and OVIII-K$\alpha$ absorption lines for observations {\bf 1x} and  {\bf 5x} 
with the RGS1 and RGS2 from XMM, respectively. 
\label{fig13}} \end{figure}

The observed changes in column density (although not very drastic) could be produced by real changes in the total column towards the source or by possible variations in the covering 
factor of  partial covering absorbers. To further investigate this, and to explore if the changes in flux might be connected with changes in covering factor (as suggested by \cite{tur08}), 
we have also looked at the two extreme XMM-Newton flux states ({\bf 1x} and {\bf 5x} ). We used two intense Oxygen absorption lines, OVIII-K$\alpha$ and OVII-K$\beta$ to track these 
possible changes. As seen in Figure \ref{fig8}, these two lines track absorption by the three components found in the soft X-ray band.  Given that these lines are saturated, their 
maximum depth should depend mainly in the fraction of the source covered by the absorber ({\it i.e.} the residual observed emission should give us directly the part of the source not 
covered). 
Therefore, changes in covering factor should produce variations in the depth of these lines. This is clearly shown in Figure \ref{fig13} (upper left panel) were we present a model assuming 
a change in the covering factor from $40\%$ to $70\%$ (as reported in \cite{tur08}).  In Figure \ref{fig13} (upper right and bottom panels) we  present the normalized data for spectra 
{\bf 1x} and {\bf 5x}, and {\bf 7c}, and {\bf 9c}. It is clear that the absorption lines do not change significantly between these states, the variations are within the error bars.  Any possible 
variations does not follow the expected direction for a change in covering factor responsible for the flux changes. Rather, the data hints to a change where the absorption lines become 
deeper during the high states, as expected if the gas is responding to the changes in the continuum. The same comparison was performed among the Chandra data with a similar results. 

\begin{figure}
\begin{center}
\begin{tabular}{cc}
\includegraphics[width=2.5in]{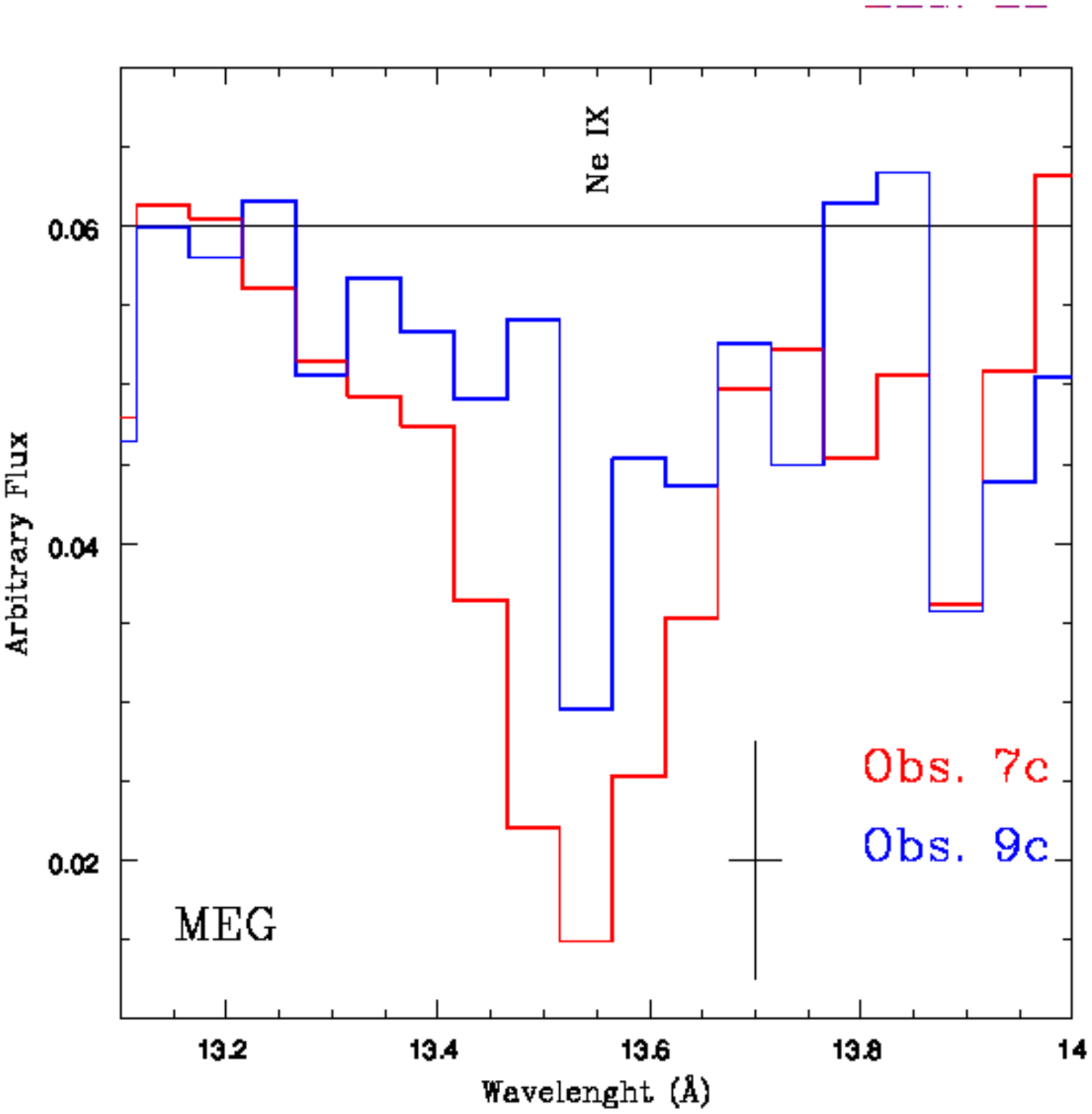}
\end{tabular}
\end{center}
\caption{ Ne IX absorption line compared between Chandra observations {\bf 7c} and {\bf 9c} in
[13, 14 $\AA$] range. 
\label{fig14}}
\end{figure}

The most striking result is obtained when comparing the NeIX absorption feature between Chandra observations. As it can be observed in Figure \ref{fig14}, there is a significant 
change in the depth of NeIX transition, between observations {\bf 7c} and {\bf 9c}. This change can be easily understood as the response in ionization state of the gas in response 
to impinging flux. Furthermore, the change is exactly the opposite to the one expected if the changes in flux were produced by changes in the covering factor. 

Therefore, the high resolution spectra do not support a scenario where the covering factor of the absorber is changing. This result is consistent with those of \cite{meh10}, who 
reached the same conclusion using data models. Thus, both data and models suggest that the changes in flux are not due to changes in covering factor of the ionized absorber as
suggested by \cite{tur08}. Rather, our analysis (using models and direct comparison of spectra) indicates that the changes in spectral features are consistent with opacity changes 
following flux variations.

\section{Discussion}

\subsection{Location and Density of the Ionization Components}

The time variability analysis of the ionized absorber can lead us to constrain the number density of the gas and thus break the degeneracy between this quantity and the distance to 
the continuum source. As shown by \cite{nic99}, the photoionization equilibrium time of a cloud of gas in response to variations in the ionized continuum is inversely proportional to the 
density of the cloud, according to the following formula:

$$t_{eq}^{x^{i},x^{i+1}} \sim \Big\{ \frac {1}{\alpha_{rec}(x^{i},T_{e})_{eq}n_{e}} \Big\} \times   $$
$$ \Big\{ \frac {1}{[\alpha_{rec}(x^{i-1},T_{e})/\alpha_{rec}(x^{i},T_{e})]_{eq}+
(n_{X^{i+1}}/n_{X^{i}})] }\Big\} $$

where t$_{eq}$ is the photoionization equilibrium time, T$_{e}$ is the electron temperature and $\alpha_{rec}(x^{i})$ is the radiative recombination rate coefficient (obtained from 
\cite{shu82}) for the ions $x^{i}$ and $x^{i+1}$). We have used the response times detected in \S \ref{variability} to constrain in the photoionization equilibrium timescale of the gas, 
and thus, in the density of the gas.

\subsubsection{Phase MI Location}

The ionization state of this component follows closely the flux changes. Thus, the photoionization equilibrium timescale must be smaller than the typical elapsed time between 
observations. Using observations {\bf 4c} to {\bf 5x}, we find t$_{1}$=15.12 ksec (t$_{1}$ is defined as the time between observations {\bf 4c} to {\bf 5x}), thus t$_{eq} \lesssim$ 
t$_{1}$ = 15.12 ksec. 

Using the electronic temperature for this phase (T$_{e}=9.8\times10^{4}$ K) and the recombination rates for OVII and OVIII, we find n$_{e}> 4.9\times10^{6}$ $cm^{-3}$. From the 
measured value of U $^{1}$ and the inferred luminosity of photons of the central engine ($Q\cong4.0\times10^{53}$ photons/s), this implies that the distance of component {\bf MI} to 
the central source must be R$_{MI} < 3.4 \times 10^{17}$ cm or R$_{MI} < 0.11$ pc. \cite{kra02} estimated the dust sublimation radius of NGC 3516 as R$_{sub} \lesssim 2.7\times
10^{17}$ cm, assuming a dust sublimation temperature of 1500 K. Thus, this component is marginally consistent with the dust sublimation radius. However, we note that in the torus 
wind model of \cite{kro01}, the  outflow  must arise far beyond this radius, well into the ``obscuring torus''. (see \cite{kro07} for a detailed discussion). Then, we consider more likely 
that this component arises from the accretion disk as suggested by \cite{net02}. 

\subsubsection{Phase LI Location}

According with the response observed in the ionization degree of the phase {\bf LI}, two different timescales can be estimated. First, during the recombination phase from observations
{\bf 2x} and {\bf 4c} this phase does not reach photoionization equilibrium with the impinging flux (see Figure \ref{fig9}). Therefore, t$_{eq} > 125$ ksec, the time elapsed between these 
two observations. On the other hand, during the ionization episode between observations {\bf 5x} and {\bf 6c}, that lasted $\sim 50$ ksec, this absorbing component reached 
photoionization equilibrium with the source. We note that the different photoionization equilibrium timescales presented above are not incompatible with each other. Equilibration times 
are much larger during recombination phases than ionization ones. With these two constraints, and the electron temperature of this component  T$_{e,LI}=3.0\times10^{4}$ K, 
we find a density $7.8\times10^{5}$ cm$^{-3}$ $\lesssim n_{e,LI} \lesssim 1.9\times10^{6}$ cm$^{-3}$. This in turn, imply that the distance of this component to the central source 
should be between $1.8\times10^{18}$ cm $\lesssim R_{LI} \lesssim 2.9 \times 10^{18}$ cm. This component is likely arising from the dusty torus, as suggested by \cite{kro01}. 
However, we note that the flow might originate close to the region were we are observing it. Thus, a disk wind cannot be strictly ruled out.

\subsubsection{Phase HI Location}

The location of phase {\bf HI} is more uncertain. The response timescale to the changes in the ionization flux should be greater than the duration of the low flux episode in the entire 
observation. From Figure \ref{fig9}, the low flux episode lasted $\sim$ 200 ksec from observations {\bf 3c} to {\bf 6c}. Thus, the density for phase {\bf HI} must be n$_{e} > 6.4\times 
10^{4}$ $cm^{-3}$ (using a temperature of T$_{e,HI}=4.5\times 10^{5}$ K and FeXVIII and Fe XIX charge states). Therefore, the minimum distance to the central source for phase 
{\bf HI} is R$_{HI}\sim 4.0\times10^{17}$ cm, which is around $\sim$ 0.13 pc. 
The location of this component is consistent with a flow arising in the dusty torus, or a flow arising from the accretion disk, observed further out.  

As explained in \S \ref {vh}, the lack of variation in component VH cannot be used to obtain a reliable constrain on distance, given that at higher energies the flux variations are much 
smaller. 

\subsection{Does the absorber forms a multi-phase medium?\label{multiph}.}

It has been suggested by several studies (e.g. Krongold et al. (2003, 2005); \cite{car09}) that the structure of ionized absorbers could be that of a multi-phase medium in pressure 
balance. A multi-phase medium requires absorbers with similar outflow velocities to avoid drag forces to destroy one of the components. Clearly, component {\bf MI} cannot be part of 
the same flow as components {\bf LI} and {\bf HI}, as these components are located farther out from the central source and they have a much faster outflow velocity. However, 
components {\bf LI} and {\bf HI} have a similar outflow velocity and they may be located at a similar distance from the central source. Then, it is possible that this components indeed 
form a multi-phase medium. 

\begin{figure}
   \centering
   \includegraphics[width=1\linewidth]{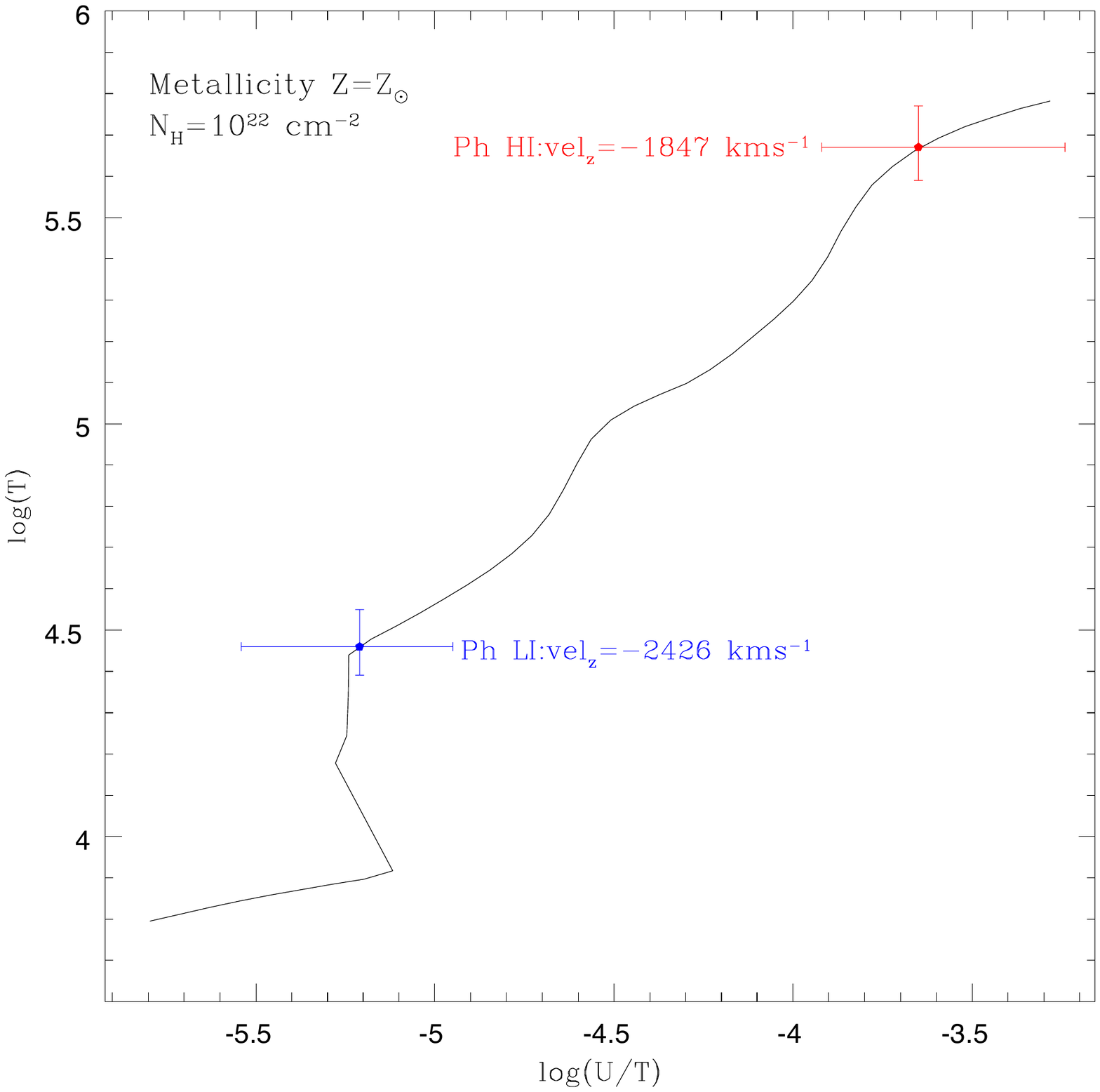}
   \caption{Thermal stability curve {\it S} of NGC 3516 SED supposing log N$_{H}=10^{22}$ cm$^{-2}$
   and a solar metallicity Z$_{\odot}$. Over-plotted there are the ionized absorbers {\bf HI}
   and {\bf LI}. 
   \label{fig15}}
\end{figure}

In order to further study this possibility, the thermal equilibrium ({\it S}-curve) obtained for the spectral distribution (SED) in this analysis is plotted in Figure \ref{fig15}. This curve 
presents the points where the gas temperature (logT) and pressure lie in thermal equilibrium. The gas pressure is inversely proportional to the quantity log(U/T), which can be 
directly derived in our analysis. Thus, two components can be part of a multi-phase flow if they have different temperatures but similar log(U/T) values. Assuming solar metallicities
(Z$_{\odot}$) the {\it S}-curve for NGC 3516 (Figure \ref{fig15}) is not multi-valued. Thus, it is not possible that these two components exist in pressure equilibrium.

\begin{figure}
   \centering
   \includegraphics[width=1\linewidth]{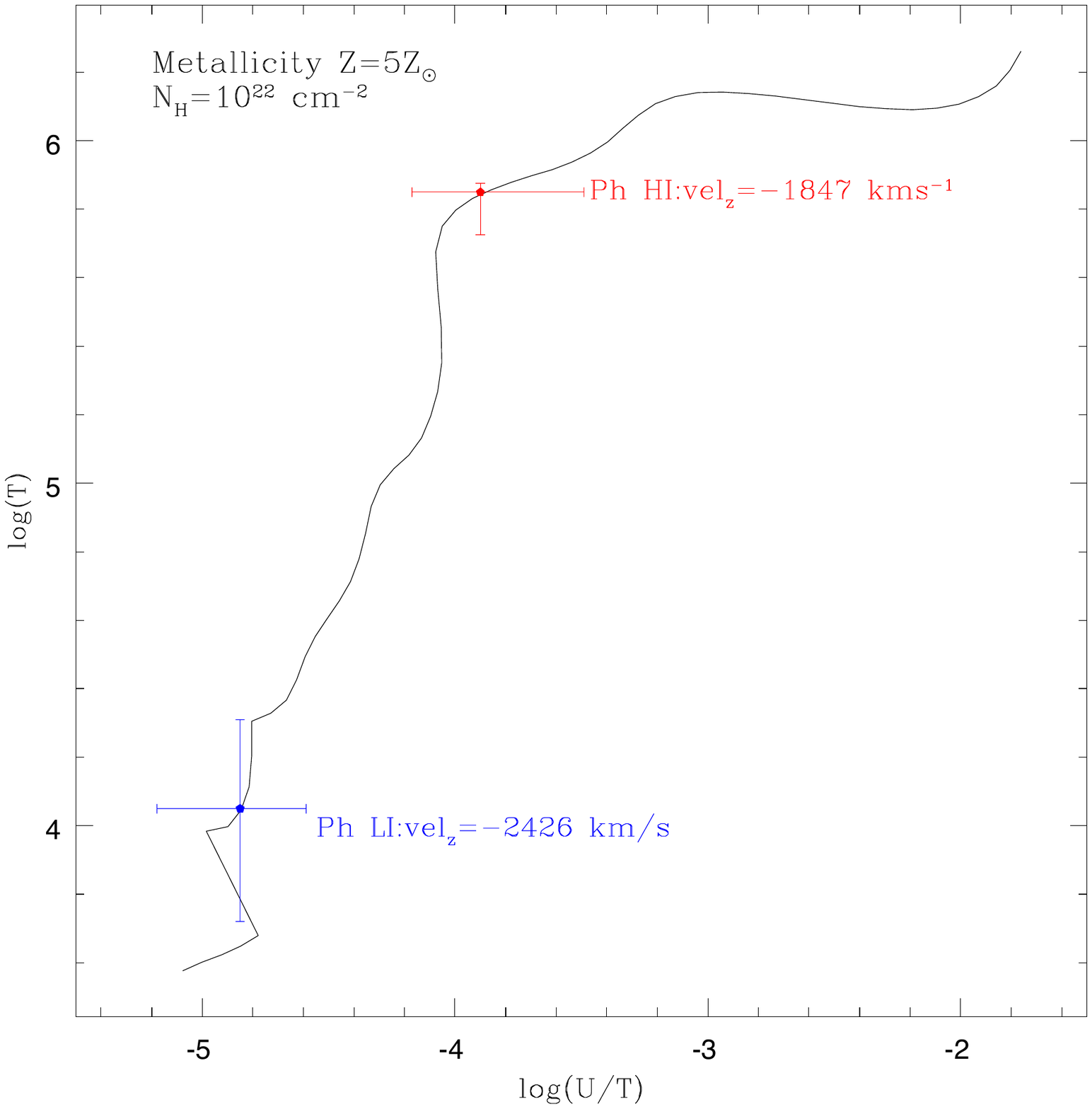}
   \caption{Thermal stability {\it S}-curve of NGC 3516 SED supposing log N$_{H}=10^{22}$ cm$^{-2}$
   and supra-solar metallicity 5Z$_{\odot}$. Over-plotted there are the ionized absorbers {\bf HI} and {\bf LI}.
   \label{fig16}}
\end{figure}

However, the shape of the {\it S}-curve results very sensitive to external heating and cooling sources. Additional cooling can be produced by material with metallicities higher than the 
solar. In fact, in AGN, supra-solar metallicities are expected (e.g. \cite{mih83} \& \cite{fie07}). 
A higher amount of cooling material creates an {\it S}-curve more vertical and together a possible ``multi-phase'' system. With the goal to explore this possibility we present a model 
with 5 times solar metallicity (5Z$_{\odot}$). Figure \ref{fig16} shows that if  the metallicity of the WA is higher than solar, then components {\bf LI} and {\bf HI}, could coexist in 
pressure equilibrium given that the pressure of both absorbers are in the same order of magnitude. As seen in the plot, for a metallicity  5Z$_{\odot}$, the difference between the 
gas pressure of these components is not significative. 

\subsection{Intrinsic Variability vs. Obscuration Scenarios for the Flux changes in NGC 3516}

Two possible scenarios have been suggested to explain the observed variability in the X-ray emission of NGC 3516. The first one suggests intrinsic variations of the central source 
\citep{meh10}. In the second scenario, explored by \cite{tur08} and \cite{mar08}, they proposed that the variations are due to obscuration of the central source produced by near blobs 
of material, either ionized or neutral. In particular \cite{tur08} suggest that an absorbing component of ionized material (their "heavy absorber") presents a change in the covering factor, 
that is responsible for the observed flux variations. Such component would produce absorption by ions of OVIII and OVII \citep{tur08}. However, we could not find evidence of 
covering factor variations in the OVII-K$\beta$ and OVIII-K$\alpha$ lines (Figure \ref{fig13}).

To further explored this possibility, we fit the spectra using a partial covering absorber. We note that partial covering is not required by the data, as a fit including it does not improve 
over a fit with fully covering absorbers (in agreement with \cite{meh10}). Nevertheless, based on {\bf Model C} a partial covering was tested in all the possible combinations, including this 
factor in one, two and the three ionizing components, that imprint spectral features in the soft X-ray band. 

The best solution found included partial covering in phase {\bf HI}. However, the value of the covering factor is almost unity in all observations (see Table \ref{Table10}), and does not 
anti-correlate with the flux. An F-test indicates that varying the covering factor is not needed (confidence of 57\% if we compare the model which does not include the covering factor and 
this model). Thus, our results do not support a scenario in which the covering factor of the warm absorbers is responsible for the flux variations. If these variation are indeed produced by 
obscuring material, this material should be located closer in than the warm absorbing components, as components {\bf LI} and {\bf MI} are responding to the observed flux variations.  

\begin{deluxetable}{r|rrrrrrrrrr|} 
\tabletypesize{\scriptsize}
\tablecolumns{10} 
\tablewidth{0pc} 
\tablecaption{Covering factor values. The percentage of the covering factor presents a typical error of 5\% \label{Table10}. } 
\tablehead{ 
\colhead{\bf Obs.} & \colhead{\bf 1x} & \colhead{\bf 2x} & \colhead{\bf 3c} & \colhead{\bf 4c} 
& \colhead{\bf 5x} & \colhead{\bf 6c} & \colhead{\bf 7c} & \colhead{\bf 8x} & \colhead{\bf 9c} }
\startdata
{\bf Covering Factor \%} & $96\%$ & $89\%$ & $84\%$ & $100\%$ & $100\%$ & $93\%$ & $88\%$ & $99\%$ & $93\%$ \\ 
\tablecaption{}
\enddata  

\end{deluxetable}

\section{Conclusions}

We summarize our results as follows:

1. We modelled the nine X-ray 2006 data of XMM-Newton and Chandra. The best statistical model consists of a complex and  variable continuum emission component, and also 
includes nine emission lines of different ionized species in the soft X-ray band. An additional emission line is required in the X-ray hard band, where the Fe-K$\alpha$ emission line is 
present. The continuum emission is absorbed by four different ionized phases (warm absorbers): three of them produce features in the soft X-ray band and the fourth, with the highest 
ionization degree, only generates the Fe-K$\alpha$ absorber complex detected in the 6.7 - 7 keV range. 

2. The ionization state of the two components with the lowest ionization degree (phases {\bf MI} and {\bf LI}) is responding to the changes in the ionizing continuum. The other two 
absorbing components (phases {\bf VH} and {\bf HI}) do not seem to react to the ionizing continuum variations.

3. The observed variations in the Fe M-shell UTA  (produced by phases {\bf MI} and {\bf LI}) between observations {\bf 1x} and {\bf 5x}  were found to be similar to what is expected for 
ionized material close to photoionization equilibrium. 

4. The  response timescale of the phases {\bf MI} and {\bf LI} allow us to constrain their location: for phase {\bf MI} the location is  R$_{MI} < 2.7 \times 10^{17}$ cm. For component 
{\bf LI}, the location must be within the range $1.8\times10^{18}$ cm $\lesssim R_{MI} \lesssim 2.9 \times 10^{18}$ cm. For component  {\bf HI}, a lower limit to the location was 
derived: R$_{HI} \gtrsim 4.0\times10^{17}$ $cm$. The lack of variations in component {\bf VH} cannot be used to find a reliable constraint on its location (see \S \ref{vh}).

5. There is no evidence of a depth change in the main absorber transitions (OVII-K${\beta}$, OVIII-K${\alpha}$ and Ne IX), as expected in a scenario were the covering factor is 
varying in the time. 

6. Finally, if a solar metallicity is assumed (Z$_\odot$), the absorbing components cannot be in pressure equilibrium. However, with higher metallicity values (5Z$_\odot$) 
the {\it S}-curve becomes more vertical and then, it becomes possible that the  absorbers {\bf HI} and {\bf LI} form a multi-phase medium.

We thank the referee for her/his thoughtful and constructive comments that helped to improve the paper. We also thank Doron Chelouche, Maria Santos-Lleo and Ehur Behar
for their invaluable help and constructive comments. This research is based on observations  obtained with XMM-Newton, an ESA science mission with instruments and contributions 
directly funded by ESA Member States, and Chandra Space Telescope of NASA.

\acknowledgments

\appendix

\end{document}